\documentclass[preprint, 3p, twocolumn]{elsarticle}

\usepackage{lineno}
\modulolinenumbers[5]
\usepackage[hidelinks]{hyperref}

\usepackage[OT1]{fontenc}

\usepackage[noend]{algorithmic}
\usepackage{xcolor}

\usepackage{graphicx,graphics}
\usepackage{amsmath,amsthm,verbatim,amssymb,amsfonts,amscd}
\usepackage{enumitem}
\usepackage{xspace}
\usepackage{subfigure}
\usepackage{cases}
\usepackage{booktabs} 
\usepackage{makecell} 
\usepackage{url}
\usepackage{algorithm}

\graphicspath{{figures/}{exp_figures/}}

\newcommand{\myparagraph}[1]{{\vspace{0.002mm}\noindent\textbf{#1}.}}

\newcommand{\REF}[1]{{\color{red} [REF]}}

\usepackage[disable]{todonotes}

\begin{document}

\begin{frontmatter}
\title{Affinity-Aware Resource Provisioning for Long-Running Applications in Shared Clusters}

\author[leeds]{Cl\'{e}ment Mommessin\corref{dag}}
\author[leeds]{Renyu Yang\corref{dag}}
\author[leeds]{Natalia V. Shakhlevich\corref{corresponding}}
\ead{n.shakhlevich@leeds.ac.uk}
\author[leeds,alibaba]{Xiaoyang Sun}
\author[leeds]{Satish Kumar}
\author[alibaba]{Junqing Xiao}
\author[leeds]{Jie Xu\corref{corresponding}}
\ead{j.xu@leeds.ac.uk}


\cortext[dag]{Co-first authors}
\cortext[corresponding]{Corresponding authors}

\address[leeds]{School of Computing, University of Leeds, UK}
\address[alibaba]{Alibaba Group, China}

\newpage
\begin{abstract}
Resource provisioning plays a pivotal role in determining the right amount of infrastructure resource to run applications and target the global decarbonization goal.
A significant portion of production clusters is now dedicated to long-running applications (LRAs), which are typically in the form of microservices and executed in the order of hours or even months.
It is therefore practically important to plan ahead the placement of LRAs in a shared cluster so that the number of compute nodes required by them can be minimized to reduce carbon footprint and lower operational costs.
Existing works on LRA scheduling are often application-agnostic, without particularly addressing the constraining requirements imposed by LRAs, such as co-location affinity constraints and time-varying resource requirements.

In this paper, we present an affinity-aware resource provisioning approach for deploying large-scale LRAs in a shared cluster subject to multiple constraints, with the objective of minimizing the number of compute nodes in use.
We investigate a broad range of solution algorithms which fall into three main categories: Application-Centric, Node-Centric, and Multi-Node approaches, and tune them for typical large-scale real-world scenarios.
Experimental studies driven by the Alibaba Tianchi dataset show that our algorithms can achieve competitive scheduling effectiveness and running time, as compared with the heuristics used by the latest work including Medea and \textsc{Lra}Sched.
Best results are obtained by the Application-Centric algorithms, if the algorithm's running time is of primary concern, and by Multi-Node algorithms, if the solution quality is of primary concern.
\end{abstract}


\end{frontmatter}

\section{Introduction}
\label{sec:intro}

Resource provisioning in large-scale compute clusters is of the utmost importance in IT infrastructure capacity management~\cite{torres2020sre} and critical to the overall stability and performance of a cluster~\cite{cherkasova2004sla}. It must take into account the characteristics of workloads and use cases in order to correctly size a cluster and minimize the cost of workload deployment.
This is of paramount importance in providing effective pathways to global decarbonization of cloud datacenters that are among the world's biggest power consumers.

Traditional workloads in clusters are data analytic batch jobs~\cite{dean2008mapreduce, saha2015apache, zhang2014fuxi} with short-lived tasks (in the order of seconds). However, long-running applications (LRAs) -- such as latency-sensitive databases, user-facing services, streaming processing frameworks, etc. -- have now become another main type of workloads supported by production clusters (Google~\cite{verma2015large}, Microsoft~\cite{garefalakis2018medea}, Alibaba~\cite{liu2018elasticity}). In particular, across six analytics clusters at Microsoft, each comprising tens of thousands of machines, at least $10\%$ of each cluster’s machines are used for LRAs and two clusters are used exclusively for LRAs~\cite{garefalakis2018medea}. In Alibaba, $94.2\%$ of the total CPU capacity in a cluster is allocated to LRAs~\cite{guo2019limits}.   
In fact, microservice architecture has been the key enabler to build up large-scale IT infrastructures. Each individual microservice -- practically instantiated as an LRA that can be independently implemented, built and maintained -- is hosted in a long-lived container that usually executes for a long time frame (from hours to months) either for iterative computations in memory or for handling web requests. An LRA often makes use of multiple replicas of it to ensure low latency, fault tolerance, and high availability~\cite{verma2015large,ladin1992providing,k8s}.

While it is appealing to build up complex enterprise IT systems consisting of a very large number of LRAs, there are many challenges associated with co-location, LRA multiplicity and heterogeneity. 
In reservation-based infrastructure, LRAs typically need to reserve multi-dimensional resources ahead of their execution, and their resource usage usually has strong temporal patterns. To optimize the performance and resilience, an LRA has application-specific placement preferences or exclusions when it is co-located with other LRAs.
For instance, some LRAs are often required to be co-located to save network bandwidth and reduce latency or to be separately placed to reduce resource contention and performance interference.
The ever-increasing scale of the number of new LRAs to be deployed (tens of thousands) and the corresponding affinity relationships further complicate resource reservation.
In a nutshell, a robust and scalable resource provisioning scheme should tackle multi-dimensional temporal resource requests and LRA-level affinities, i.e., it should address placement of identical replicas incurred by each LRA, resolve replica conflicts stemming from the affinity constraints, and handle efficiently large-scale LRA deployment scenarios.

To the best of our knowledge, none of the existing studies to date addresses all these requirements at the same time, although every single requirement might have been considered.
Most of the existing work (e.g., \cite{zhang2014fuxi,verma2015large,vavilapalli2013apache,sun2018rose,yang2020performance}) is application-agnostic and only focuses on node-related affinity, neglecting inter-application affinity constraints. 
Kubernetes~\cite{k8s} and Medea~\cite{garefalakis2018medea} address the application-related affinity, but do not address the requirement of scheduling all LRAs as a global optimization problem:
Kubernetes schedules one LRA replica (pod) at a decision point, while Medea aims at runtime scheduling of relatively small batches of LRAs periodically.
\textsc{Lra}Sched~\cite{LRASched2021} only addresses the intra-application affinity constraints.
Additionally, the capability of handling massive-scale scheduling problems of these three solutions has not been fully investigated.

The problem we study is to minimize the number of compute nodes required for accommodating LRAs in a shared cluster, subject to a set of strict resource and affinity constraints. We formulate the problem as an ILP and develop a new system model that can be considered as a generalization of the combinatorial optimization problems of Vector Bin Packing and Bin Packing with Conflicts~\cite{BinPacking-Handbook2013}.
Considering the diversity of real-world scenarios that gives rise to instances with a variety of characteristics, a fast heuristic, successful for one scenario, may perform poorly on another.
This motivates us to develop an algorithm suite that can be used by practitioners for selecting the best performing heuristics that best fit the specific needs of a given scheduling scenario.
To illustrate the capabilities of the suite, we perform experiments on instances generated from the Alibaba Tianchi dataset~\cite{tianchidata} and compare the winning approaches from our suite with the best performing published algorithms: two heuristics from Medea~\cite{garefalakis2018medea}, namely \textit{TagPopularity} and \textit{NodeCandidate}, as well as a heuristic based on the \textit{Fitness} measure introduced in \textsc{Lra}Sched~\cite{LRASched2021}.
A high-level summary of the most successful algorithms in our toolkit and those published in the literature is presented in Fig.~\ref{fig:summary}, Section~\ref{sec:exp}.

Our suite consists of three groups of algorithms: Application-Centric, Node-Centric and the Multi-Node approaches. The first two algorithm groups stem from the state-of-the-art research on Vector Bin Packing and Bin Packing with Conflicts~\cite{BinPacking-Handbook2013}. The third algorithm group is particularly successful in the presence of LRA replicas and associated affinity restrictions.
While Application-Centric algorithms are recommended when the computation time is required to be as small as possible, the Multi-Node algorithms deliver solutions of best quality (within only $0.3\%$ deviation from the lower bound), with a larger running time.
Node-Centric Algorithms place themselves in between, offering a trade-off between solution quality and time to find a solution.

To summarize, the main contributions of this paper are as follows:
\begin{itemize}[leftmargin=1.5em]
    \item Formulating a resource provisioning problem to address temporal resource requests and application-level affinity constraints (\S\ref{sec:model});
    \item Devising an algorithm suite to provide adaptable solutions to a variety of real-world scenarios (\S\ref{sec:solution});
    \item Selecting, via extensive computational experiments, a collection of best performing algorithms that can effectively handle large-scale LRA deployment (\S\ref{sec:exp}), focused on the use-case of the Alibaba Tianchi dataset~\cite{tianchidata};
    \item Elaborating algorithm recommendations providing a trade off between computation time and solution quality when confronted with different scenarios (\S\ref{sec:choice}).
\end{itemize}

Our findings can serve as the basis for practitioners and researchers for optimizing the resource provisioning and capacity planning to handle large-scale LRA placement in different scenarios.

\section{Background and Motivation}\label{sec:background}

\subsection{Microservice and Long-running Applications}

Cloud services and enterprise IT systems have been experiencing a major shift from monolithic applications that encompass the whole functionality within a software package (e.g., the full-stack LAMP application) to thousands of loosely-coupled microservices that can be independently built and maintained.
According to Statista survey~\cite{statista}, in 2021, $85\%$ of respondents from large organizations with $5,000$ or more employees stated that they had been using microservices in their software development environments.

As a key enabler, microservice architecture is particularly supportive to build extensible and loosely-coupled systems at scale. Enterprise microservices can be considered as an important and widely popular types of long-running applications (LRAs). They are typically hosted in long-lived containers that can run for hours, or even months, and consist of a diverse mix of applications from web servers to databases. Such applications are long-standing, user-facing and interactive services, working in ``request-and-response'' manner to serve user requests.
Representative examples of LRAs include streaming processing frameworks (Storm~\cite{storm}, Flink~\cite{flink}, Kafka streams~\cite{kafkastream}),
latency-sensitive database applications (HBase~\cite{hbase} and MongoDB~\cite{mongodb}),
and data-intensive in-memory computing frameworks (Spark~\cite{zaharia2016apache}, Tensorflow~\cite{abadi2016tensorflow}). 

\subsection{Resource Provisioning}

LRAs need to be deployed into on-premises or cloud infrastructure. Resource provisioning -- one of the key elements in capacity management \cite{torres2020sre} -- plays a pivotal role in determining the \textit{initial} amount of infrastructure capacity (required resources) that can run a collection of applications. 
Particularly, for a homogeneous computing cluster where each node has the same hardware and the same operating system, infrastructure capacity can be regarded as the number of compute nodes (bare metal servers or virtual machines in a virtualized cloud cluster).

Most large-scale infrastructure managers \cite{verma2015large,liu2018elasticity,cheng2018characterizing} adopt \textit{reservation}-based resource requests and resource allocations, i.e., application users or developers are required to specify the number of resources required (CPU cores, RAM, GPUs, etc.) at the submission of the applications.
To reduce carbon footprint and lower operational costs, one simple yet prevalent task of resource provisioning is to minimize the number of nodes capable of handling the reservation requests of a given set of LRAs. The plan-ahead before LRA deployment and execution is of major importance to IT administrators to facilitate a better understanding of resource requirement and to resize the infrastructure configuration in an economical and environmental-friendly manner.

\begin{figure*}[t]
\centering
\subfigure[CPU usage]{
\includegraphics[width=0.4\textwidth]{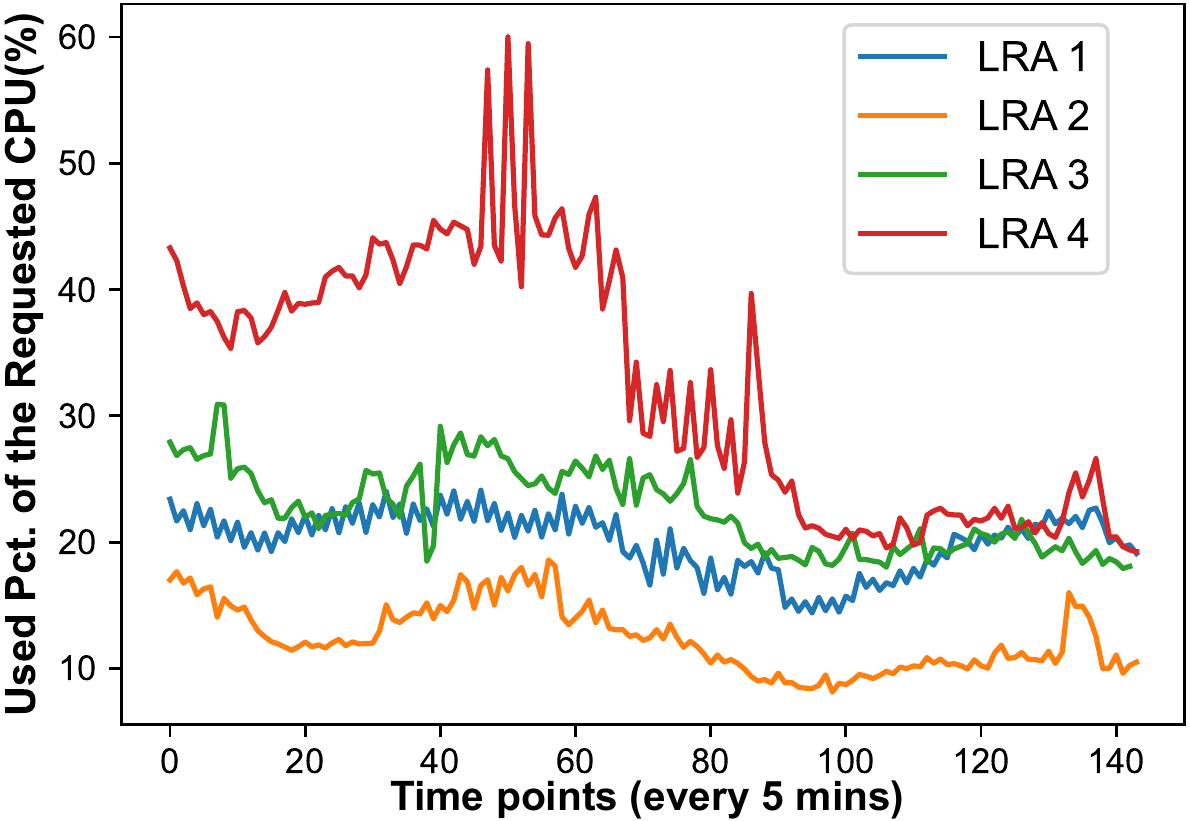}
\label{fig:eff-cap-scale:varyingresreq:a}}
\hspace{3em}
\subfigure[Memory usage]{
\includegraphics[width=0.4\textwidth]{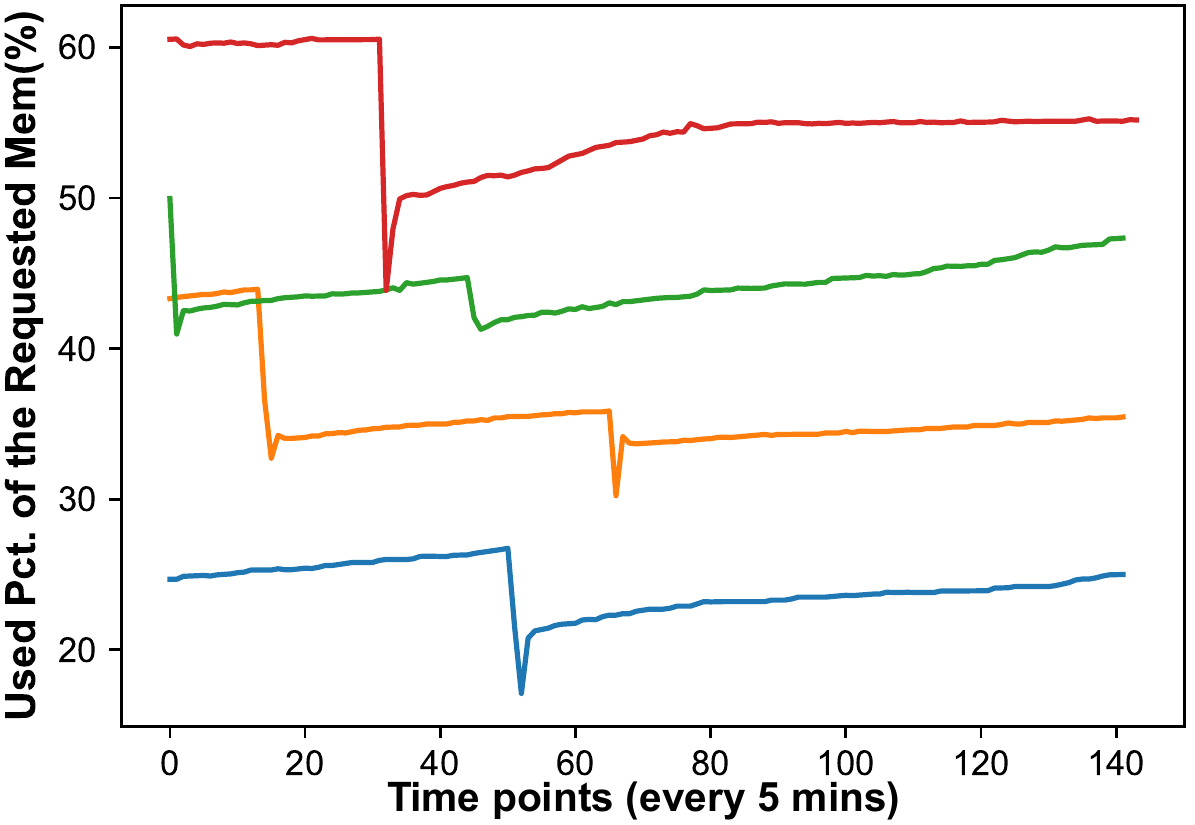}
\label{fig:eff-cap-scale:varyingresreq:b}}
\vspace{-1em}
\caption{CPU and memory temporal usage over 12 hours of four anonymous LRAs in Alibaba cluster trace}
\label{fig:eff-cap-scale}
\end{figure*}

\subsection{Problem Scope and Challenges}
\label{sec:model:goal}

While runtime LRA scheduling is well addressed by cluster schedulers~\cite{verma2015large,yang2020performance,lo2015heracles,hu2020toposch}, 
this work focuses on addressing a planning problem for resource provisioning as we envisage the importance of pre-execution planning to the cost reduction of infrastructure management.
The resource planner aims to work out the best option for deploying the LRAs ahead of their execution, given that all information of LRAs to be submitted is foreknown, to ensure a predictable LRA execution.

We highlight challenging requirements for the planning problem we address in this paper. 

\begin{itemize}[leftmargin=1.5em]
    \item \textbf{[R1] Multi-dimensional and time-varying resource requirements.}
    LRAs usually require resources of different types (CPU cores, memory, disk, etc.). Additionally, LRAs experience a noticeable temporal resource dynamicity over time.
    Fig.~\ref{fig:eff-cap-scale} illustrates the dynamicity of CPU and memory usage of multiple co-located LRAs over 12 hours, observed from the Alibaba Cluster Trace~\cite{alibabaclutertrace}.
    Such dynamicity can be captured through history-based profiling as most LRA workloads run in a recurring manner and have strong temporal pattern ~\cite{cortez2017resource}, which helps to unlock the potential of accurate requirement models and workload co-location in large-scale clusters~\cite{liu2018elasticity,guo2019limits,cheng2018characterizing}.
    The owner of an LRA typically needs to determine the resource demands (e.g., through extracting resource skyline based on the resource requirement  model) and translate the temporal requirement into resource reservation.
    
    \item \textbf{[R2] Application-level affinity constraints.} 
    Affinity constraints encompass placement preferences or exclusions between LRAs. While node-related affinity specifies which nodes an LRA is eligible to be placed on, application-level affinity specifies how many replicas of an LRA can be placed jointly given the co-located LRAs on a node.
    These constraints are completely application-specific.
    For example, data producing and data consuming applications could be co-located on the same node for sharing intermediate data to save network bandwidth and reduce network latency.
    To avoid excessive performance interference, latency-sensitive streaming applications should not be co-located on the same node.
    However, for some LRAs, it is reasonable to co-locate their replicas within the same available zone, which would help to ease service management, reduce the cost of synchronization or data communication between applications.
    Running applications without satisfying such constraints would lead to unexpected application slowdown or system turbulence.
    Such affinity requirements are usually specified in the configuration (e.g., in a YAML/JSON file) to flag LRA-specific performance preferences and QoS requirements, before the deployment requests are submitted to the infrastructure manager.
    
    \item \textbf{[R3] Large-scale LRA deployment}.
    Launching tens of thousands of LRAs has now become the norm rather than the exception for cloud service providers in the face of new cluster initialization.
    This increases the management complexity of deploying large-scale LRAs.
    Each LRA has its own specific deployment and resource requirements (e.g., CPU cores, RAM and persistent storage).
    Therefore, the infrastructure manager needs to be robust and scalable enough to make (near-)optimal decisions, incorporating in the planning a huge number of resource and affinity requirements, in the initial deployment stage. 
\end{itemize}

The existing works only partially solve the above research challenges.
Unlike runtime LRA scheduling, that aims to achieve low scheduling latency (in the order of seconds or milliseconds), the main task of pre-execution planning is to precisely place the LRAs and to determine the amount of required resources in the IT infrastructure while satisfying all sophisticated specific constraints of applications.

Obviously, for the resource planner, it is worth trading the planning time for solution quality. This trade-off in the planning procedure is particularly pivotal as low-quality LRA placement may incur excessive cost in LRA re-scheduling and container migration, which is expensive due to the huge amount of state and disk data to migrate over the network and unacceptable service downtime.
We believe an optimization-based plan-ahead is a necessary and promising means for effective resource provisioning.
Our work aims at integrating the above requirements into a holistic system model, and developing a suite of algorithms able to solve the resource provisioning problem and adapt to different scenarios.

\section{System Model and Problem Formulation}\label{sec:model}

\subsection{System Model} \label{sec:model:model}

Our system consists of {\it compute nodes}, which form the set $\mathcal{N}$, and {\it LRAs}, which form the set $\mathcal{L}$. Additionally, there are  {\it affinity restrictions} for some pairs of LRAs from $\mathcal{L}$.

\textbf{Compute nodes} are identical and their resources are characterized by $d$ dimensions. 
In our {\it basic} model, there are two types of resources, the number of CPU cores $C_1$ and the number of units of memory $C_2$.
It can be extended to take into account such characteristics as the size of disk storage or Last-Level Cache, the memory bandwidth, the number of GPU, etc. 
In general, according to {\bf [R1]} of the model, a node has $d$ dimensions, with resource capacities $C_1, C_2, \ldots, C_d$. 

\textbf{LRAs} differ in a number of parameters.
In accordance with {\bf [R2]}, each LRA consists of a given number of replicas that run from time $0$ to infinity (or to a given time limit common for all LRAs).
An LRA $i \in \mathcal{L}$ has a given size $s_{ih}$ (i.e., resource requirement) in dimension $h$, $1 \le h \le d$, and that value is the same for all replicas of that LRA. For example, for the basic model, $s_{i1}$ and $s_{i2}$ are the number of CPU cores and the number of units of memory needed by each replica of LRA $i$.

In the {\it basic} model, we assume that the sizes of LRAs do not change over time.
If several replicas of LRAs are allocated to the same node, then the total size of allocated replicas in each dimension cannot exceed the node capacity in that dimension.
Thus $d$ capacity constraints should be satisfied for each node. 

In the {\it enhanced} model, the profiles of LRAs may change over time. They are approximated via piece-wise constant functions.
If the timeline is split into $T$ epochs (for example, $T$ unit-time intervals), so that within one epoch resource requirements of LRAs do not change, then the original $d$-dimensional problem, with $d$ resource types, is converted into the problem of $d'$ dimensions:
\begin{equation*}
    d'=T \times d.
\end{equation*}

\begin{figure}[t]
  \begin{center}
    \includegraphics[width = 0.48\textwidth]{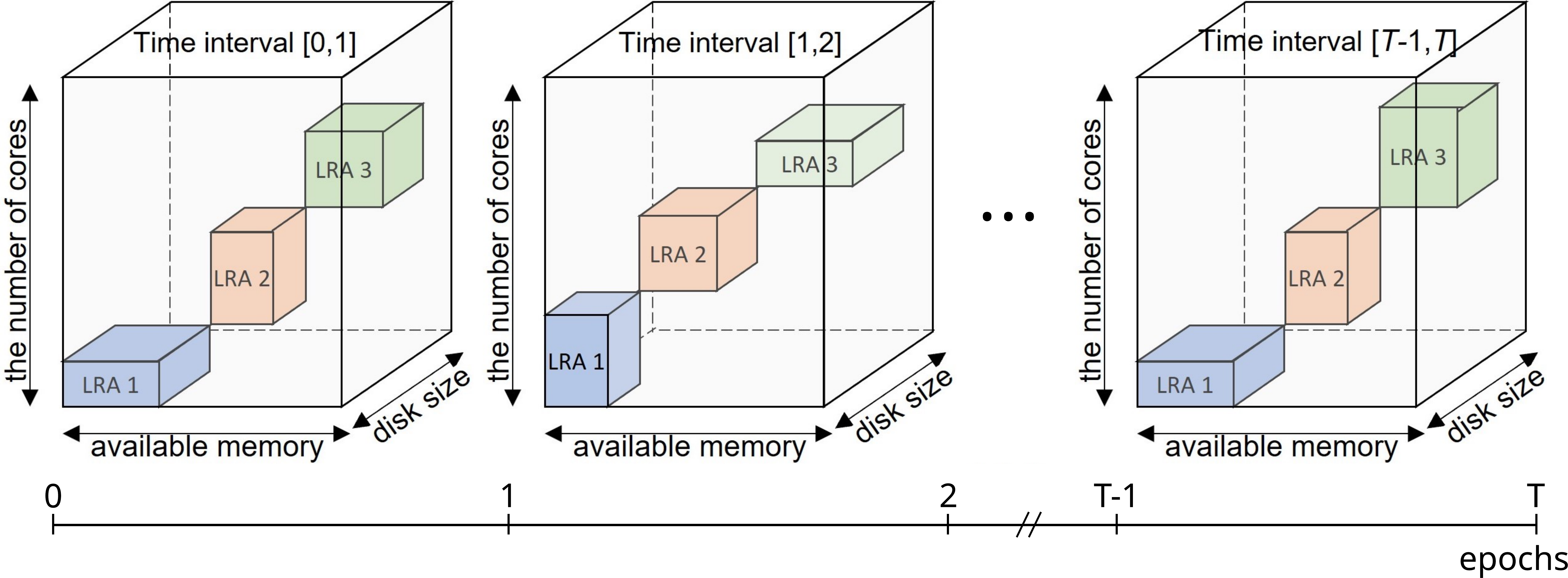}
  \end{center}
  \vspace{-1.4em}
  \caption{Allocation of three-dimensional LRAs to one node taking into account changing resource requirements over $T$ epochs}
  \label{fig:Fig-TimeVarying}
\end{figure}

Fig.~\ref{fig:Fig-TimeVarying} illustrates allocation of three LRAs to one compute node. Each LRA has specific resource requirements for $d=3$ resource types: memory, CPU and disc space.
If application requirements are static, then it is sufficient to consider only one fragment of Fig.~\ref{fig:Fig-TimeVarying}: one three-dimensional cube for a node, with LRAs placed inside it without overlaps in each dimension. If application requirements change in $T$ time intervals, then the memory, CPU and disk constraints should be considered for each time interval. For the instance considered in Fig.~\ref{fig:Fig-TimeVarying}, there is one node and $T$ snapshots of that node, with the same three LRAs allocated to the node in each of the $T$ snapshots. The resource requirements of the LRAs change, but the overall capacity of the node is not exceeded.

\textbf{Affinity restrictions} are defined for pairs of LRAs which replicas can be jointly co-located to the same node, but with some limits, or for pairs of incompatible LRAs, which cannot be co-located.
If LRA $i$ is restrictive to LRA $j$, then there is an integer \textit{affinity value} $a_{ij}$ which sets up an upper bound on the maximum number of replicas of $j$ that can be co-located on a node where at least one replica of $i$ is allocated.
Thus {\bf [R2]} of the model is characterized by the set of affinity restrictions, represented as a directed graph where vertices correspond to LRAs and arcs $(i,j)$ correspond to affinity restrictions associated with the values $a_{ij}$.

\subsection{Problem Formulation} 
\label{sec:model:form}

In a feasible solution to the resource provisioning problem, all replicas of all LRAs in the given set $\mathcal{L}$ should be allocated to a subset of $\mathcal{N}$, without violating affinity restrictions and node capacities in each of the $d$ dimensions (or, in general, $d'$ dimensions). 
The objective is to minimize the total number of nodes in use.  

We introduce an Integer Linear Programming (ILP) formulation for the resource provisioning problem with constant resource demands.
Recall that for the time-varying resource demands, the $d$-dimensional problem is converted into $d'$-dimensional problem, $d'=Td$, which implies that $d$ is replaced by $d'$ in the ILP formulation.

We use the following notations:
\begin{tabbing}
$\mathcal{L}$~~ \= for the set of LRAs, \\ 
$\mathcal{R}_i$ \> for the set of replicas of an application $i \in \mathcal{L}$, \\
$\mathcal{N}$ \> for the set of nodes, \\
$\mathcal{A}$ \> for the set of pairs $(i,j)$ of applications which\\
\> have affinity restrictions $a_{ij}$, \\
$s_{ih}$ \> for the size of resource $h$ required by a replica\\
\> of application $i \in \mathcal{L}$, in dimension $h$, $1 \le h \le d$,\\
$C_h$ \> for the capacity of each node in dimension $h$,\\
\> $1 \leq h \leq d$,\\
$a_{ij}$ \> for the affinity restriction imposed by\\
\>  application $i$ (how many  replicas of $j$ can be\\
\>  co-located together with a replica of $i$).\\
\end{tabbing}

The decision variables take $0$ -- $1$ values: 
\begin{tabbing}
$x_{irn}$ \= is equal to $1$ if the $r$th replica of application $i$ \\ 
\> is allocated to node $n$,\\
$y_{n}$ \> is equal to $1$ if node $n$ is activated and \\
\> accommodates some replica(s),\\
$z_{in}$ \> is equal to $1$ if at least one replica of \\
\> application $i$ is allocated to node $n$. 
\end{tabbing}

Additionally, we compute constants $\nu_{i}$ for the maximum number of replicas of application $i$ which can be allocated to one node, regardless of affinity restrictions from other applications:
\begin{equation} 
\label{nu}
\nu_{i}= \min \left\{ \min_{1\leq h\leq d} \left\{ \left\lfloor \frac{C_{h}}{s_{ih}}\right\rfloor \right\}, |\mathcal{R}_i| \right\} .
\end{equation}
Here $|\mathcal{R}_i|$ is the total number of replicas of application $i$ and $\lfloor C_{h}/s_{ih} \rfloor$ is the limitation associated with dimension $h$ if replicas of application $i$ are allocated to a node.
For example, in the basic model with two resource types per node, the ratios  $\lfloor C_1 / s_{i1} \rfloor$ and $\lfloor C_2 / s_{i2} \rfloor$ are related to the CPU and  memory limitations for replicas of LRA $i$.
In the enhanced model with time-varying profiles, each dimension $1 \leq h \leq Td$ gives rise to a resource restriction in the corresponding epoch.

The problem of allocating the replicas of all LRAs to the minimum number of compute nodes without exceeding node capacities and violating affinity restrictions of LRAs is modelled as the following ILP: 
\begin{subequations}
\begin{align}
\min \quad & \sum_{n \in \mathcal{N}} y_n \label{C0}\\
\text{s.t.} \quad & \sum_{n \in \mathcal{N}} x_{irn}=1, \qquad~ i \in \mathcal{L}, r \in \mathcal{R}_i, \label{C1}\\
 & \sum_{i \in \mathcal{L}} s_{ih} \sum_{r \in \mathcal{R}_i} x_{irn} \le C_{h} y_n,  \notag\\
 & \qquad\qquad\qquad\qquad~ n \in \mathcal{N}, 1 \le h\le d, \label{C2} \\
 & \sum_{r \in \mathcal{R}_i} x_{irn}   \le \nu_{i} z_{in},~ i \in \mathcal{L}, n \in \mathcal{N},\label{C3}\\
 & z_{in} \le \sum_{r \in \mathcal{R}_i} x_{irn}, \quad~ i \in \mathcal{L}, n \in \mathcal{N}, \label{C4}\\
 & \sum_{r \in \mathcal{R}_j} x_{jrn} \le a_{ij}z_{in} + \nu_{j}(1-z_{in}), \notag\\
 & \qquad\qquad\qquad\qquad~ (i,j) \in \mathcal{A}, n \in \mathcal{N}, \label{C5}\\
 & x_{irn}, ~y_n, ~z_{in} \in \{ 0, 1\},  \notag\\
 & \qquad\qquad\qquad~ i \in \mathcal{L}, r \in \mathcal{R}_i, n \in \mathcal{N}. \label{C6}
\end{align}
\end{subequations}

\noindent Objective function (\ref{C0}) is the total number of activated nodes.
Constraint~(\ref{C1}) ensures that all replicas of all applications are allocated, while constraint~(\ref{C2}) ensures that the capacity of each node is not exceeded in each dimension. The variables $y_{n}$ and $z_{in}$ are linked to $x_{irn}$ by (\ref{C2})-(\ref{C4}), and constraint (\ref{C5}) guarantees that affinity restrictions are observed.

The resource provisioning problem is NP-hard, as it generalizes the combinatorial optimization problems of \textit{Vector Bin Packing} and \textit{Bin Packing with Conflicts}~\cite{BinPacking-Handbook2013}: the Vector Bin Packing problem occurs when each LRA consists of a single replica and there are no affinity restrictions; the Bin Packing with Conflicts occurs when $d=1$, each LRA consists of a single replica, and affinity values $a_{ij}$ between two conflicting LRAs are restricted to $0$. 

As we will discuss in \S\ref{sec:exp:setting}, the presented  ILP is capable of solving medium size instances, with up to $2,000$ two-dimensional LRAs. In what follows we elaborate a broad range of heuristic methods capable of solving effectively and efficiently LRA scheduling problems typical for real-world  massive-scale systems.

\section{Our Algorithm Suite}
\label{sec:solution}
This section presents an overview of the approaches (\S\ref{sec:solution:overview}) and implementation details for a range of algorithms (\S\ref{sec:solution:app-centric} to \S\ref{ssec:MultiNode}).
We then discuss the worst-case time complexities of the algorithms (\S\ref{ssec:Tuning}) and how the algorithms published in the literature and adopted by the existing schedulers fit into our framework (\S\ref{sec:solution:sota}). 

\subsection{Overview} 
\label{sec:solution:overview}

The heuristics described in this section stem from the vast body of research on the Bin Packing Problem and its enhanced versions.
The methods distinguish in how they order the set of applications $\mathcal{L}$, the set of nodes $\mathcal{N}$ or the set of application-node pairs.
The choice of the most promising prioritization rules depends on the scenarios to which the method is applied and on the datasets.

All methods consider only {\it feasible} allocations of application replicas to the nodes, so that the node capacities are not exceeded for each of the $d$ resource types in each of the $T$ epochs, and the affinity restrictions for the applications already allocated are observed. By allocating replicas to the nodes in a {\it feasible} fashion we guarantee that requirements {\bf [R1]-[R2]} are satisfied. To handle requirement {\bf [R3]} we strive to achieve fast running times for our heuristics. \\

\myparagraph{Application-Centric approach} This approach considers the applications one by one in accordance with their ordering in  $\mathcal{L}$. For a current application, it selects the first {\it feasible} node in the ordered list $\mathcal{N}$ and allocates the maximum number of replicas to that node. It then selects the next {\it feasible} node from $\mathcal{N}$ to continue allocation of the replicas of the current application. After all replicas of the current application are allocated, the algorithm proceeds with the next application in $\mathcal{L}$, etc.
The rules for ordering $\mathcal{L}$ and $\mathcal{N}$ are formulated in \S\ref{sec:solution:app-centric}, using the state-of-the-art findings in the body of research on Bin Packing and Vector Bin Packing~\cite{BinPacking-Handbook2013,MartelloToth-book-Ch8,Panigrahy}.
Algorithm~\ref{alg:app-centric} outlines the pseudo-code of this approach.\\

\begin{algorithm}[t]
\caption{Application-Centric approach}\label{alg:app-centric}
\begin{algorithmic}[1]
\STATE{Activate node $n=1$ and set $\mathcal{N} \leftarrow \{1\}$}
\WHILE{there are unallocated LRAs}
\STATE{Select $i\in \mathcal{L}$ using a predefined rule}
\WHILE{not all replicas of $i$ are allocated}
\IF{no node from $\mathcal{N}$ can accommodate $i$}
\STATE{Set  $n \leftarrow n+1$, $\mathcal{N} \leftarrow \mathcal{N}\cup \{n\}$ and activate node $n$}
\ENDIF
\STATE{Select $n^* \in \mathcal{N}$, feasible for $i$, using a predefined rule}
\STATE{Allocate the maximum number of replicas of $i$ to $n^*$}
\ENDWHILE
\STATE{Remove $i$ from $\mathcal{L}$}
\ENDWHILE
\end{algorithmic}
\end{algorithm}

\myparagraph{Node-Centric approach} This approach considers the nodes one by one in accordance with their numbering in list $\mathcal{N}$.
For a current node, the algorithm selects from the list of non-allocated applications the one which is {\it feasible} for the current node and has the largest {\it application-node score}.
The maximum number of replicas of that application are allocated to the node.
If the node is not fully packed, then the {\it application-node scores} are recalculated, taking into account the residual capacity of the current node, and the application delivering the highest score is used for loading the node.
The process continues until no feasible application for the current node can be found on the list $\mathcal{L}$.
The algorithm then proceeds with the next node in the list, etc. The scoring rules are formulated in \S\ref{sec:solution:node-centric}, using the findings in the body of research on the Vector Bin Packing problem~\cite{Panigrahy}.
Algorithm~\ref{alg:node-centric} outlines the pseudo-code of this approach.\\

\begin{algorithm}[t]
\caption{Node-Centric approach}\label{alg:node-centric}
\begin{algorithmic}[1] 
\STATE{Activate node $n=1$ and set $\mathcal{N} \leftarrow  \{1\}$}
\WHILE{there are unallocated LRAs}
\IF{no $i\in \mathcal{L}$  is feasible for $n$}
\STATE{Set $n \leftarrow n+1$, $\mathcal{N} \leftarrow \mathcal{N}\cup \{n\}$ and activate node $n$}
\ENDIF
\STATE{Select $i^*\in \mathcal{L}$ which is feasible for $n$ and delivers the maximum score}
\STATE{Allocate the maximum number of replicas of $i^*$ to $n$}
\IF{all replicas of $i^*$ are allocated} 
\STATE{Remove $i^*$ from $\mathcal{L}$}
\ENDIF
\ENDWHILE
\end{algorithmic}
\end{algorithm}

\myparagraph{Multi-Node approach} This approach aims to overcome the myopic nature of the Application-Centric and Node-Centric algorithms. A large set of nodes is activated directly at start, and best allocation options are selected across the whole pool of  nodes. The algorithm either finds a feasible solution or declares a failure, if the number of activated nodes is too small to accommodate all applications.
The proposed approach requires that the desired number of nodes is specified as part of the input. The search for a feasible solution with the minimum number of nodes is arranged by calling the algorithm repeatedly with different trial values for the number of nodes, either via binary search, or with the trial value decremented in steps.

We distinguish between the following two version of the Multi-Node approach, whose pseudo-codes are given as Algorithms \ref{alg:spread} and \ref{alg:matching}.
\begin{itemize}[leftmargin=1.5em]
    \item The \textbf{Multi-Node approach with Replica Spreading} is the adaptation of the Application-Centric approach.
    LRAs are also considered one by one, but instead of allocating the {\it maximum} number of replicas of a current application to the highest priority node, only one replica is allocated.
    Node priorities are updated every time a replica is allocated, increasing the chances that replicas of one LRA are assigned to different nodes -- a major point of difference compared to the standard Application-Centric method.
    With replicas spread over a large pool of activated nodes, there is more flexibility for selecting compatible LRAs for future co-location: affinity constraints $a_{ij}$ become less restrictive if a small number of replicas of $i$ and $j$ are allocated to the same node. 
    
    \item The \textbf{Multi-Node approach with Application-Node Matching} is the adaptation of the Node-Centric approach.
    At each step the score for every \textit{feasible} application-node pair is computed, and the pair with the highest score is selected for extending a partial solution.
    One replica of the selected application is allocated to the corresponding node, and the scores are recalculated to define the next most promising application-node pair.
    This approach is more flexible than the original Node-Centric approach: it benefits from a larger freedom for selecting the most promising application-node pairs, with potentially better utilized resources as a result.
\end{itemize}

\begin{algorithm}[t]
\caption{Multi-Node approach with Replica Spreading}\label{alg:spread}
\begin{algorithmic}[1]
\STATE{For a given $n$, activate nodes $\mathcal{N}=\{1,2,\ldots ,n\}$\vspace{-.4cm}} 
\WHILE{there are unallocated LRAs}
\STATE{Select $i\in \mathcal{L}$ using a predefined rule}
\WHILE{not all replicas of $i$ are allocated}
\IF{no node from $\mathcal{N}$ can accommodate $i$}
\STATE{declare a failure and \textbf{break}} 
\ENDIF
\STATE{Select $n^* \in \mathcal{N}$, feasible for $i$, using a predefined rule}
\STATE{Allocate one replica of $i$ to $n^*$}
\ENDWHILE
\STATE{Remove $i$ from $\mathcal{L}$}
\ENDWHILE
\end{algorithmic}
\end{algorithm}

\begin{algorithm}[t]
\caption{Multi-Node approach with Application-Node Matching}\label{alg:matching}
\begin{algorithmic}[1]
\STATE{For a given $n$, activate nodes $\mathcal{N}=\{1,2,\ldots ,n\}$\vspace{-.4cm}} 
\WHILE{there are unallocated LRAs}
\IF{no pair $(i,n)$ is feasible ($i \in \mathcal{L}$, $n \in \mathcal{N}$)}
\STATE{declare a failure and \textbf{break}}
\ENDIF
\STATE{Select a feasible pair $(i^*,n^*)$ which delivers the maximum score}
\STATE{Allocate one replica of $i^*$ to $n^*$}
\IF{all replicas of $i^*$ are allocated} 
\STATE{Remove $i^*$ from $\mathcal{L}$}
\ENDIF
\ENDWHILE
\end{algorithmic}
\end{algorithm}

\subsection{Application-Centric Algorithms}
\label{sec:solution:app-centric}

At the core of the Application-Centric algorithms are the priority rules for ordering the list of applications $\mathcal{L}$ and the list of nodes $\mathcal{N}$.
Based on the best performing algorithms known for Bin Packing, there are three widely accepted possible orderings for the nodes $\mathcal{N}$ and two orderings for the applications $\mathcal{L}$. 

For $\mathcal{N}$, the nodes can be considered (a) in the activation order, (b) in the increasing order of a priority index, or (c) in the decreasing order of a priority index.
For $\mathcal{L}$, the applications can be considered (1) in the order of their numbering, or (2) in the decreasing order of a priority index.
The priority indices can be defined in multiple ways for the multi-dimensional problem.
In the remainder of this section, we describe the rules for calculating the priority indices of applications, denoted by \textit{size measures} and used for ordering (2) of list $\mathcal{L}$, and the rules for calculating the priority indices of nodes, denoted by \textit{residual capacity measures} and used for ordering (b) or (c) of list $\mathcal{N}$.

Depending on how the rules for $\mathcal{N}$ are combined with the rules for $\mathcal{L}$, the resulting algorithms are classified as (1a) First Fit (FF), (1b) Best Fit (BF), (1c) Worst Fit (WF), (2a) First Fit Decreasing (FFD), (2b) Best Fit Decreasing (BFD) and (2c) Worst Fit Decreasing (WFD).\\

\myparagraph{Applications' priority indices}
In the presence of resource requirement in multiple dimensions, one most significant dimension can be used for prioritizing the applications.
If no dominant dimension exists, as in the case of the Alibaba Tianchi dataset~\cite{tianchidata}, there is a need to compute a {\it combined size measure} $s_i$ for each application $i \in \mathcal{L}$ and to use it as a priority index.
Introducing a single measure allows us to address efficiently the issues related to requirement {\bf [R1]}. 

When dealing with non-comparable sizes $s_{ih}$ of LRAs, such as the number of CPU cores and memory, the values should be normalized to satisfy $s'_{ih} \in [0,1]$, which is achieved by setting $s'_{ih}=\frac{s_{ih}}{C_h}$ in each dimension $h$, $1 \leq h \leq d$.
With the normalized sizes $s'_{ih}$ of LRAs, the node capacities are set to $C'_h =1$.
In what follows, we assume that the preprocessing has been done and the normalized values are calculated.
For simplicity, we drop the prime in the notation.

The two natural combined measures are {\it Average} and {\it Max}, whose corresponding expressions are stated in the first two lines of Table~\ref{tab:CombinedSizeMeasures}. 

The remaining measures use the following notations: 
\begin{tabbing}
$W_{h}=\sum_{i\in \mathcal{L}}\left\vert R_{i}\right\vert s_{ih}$~ \= for the total demand of all \\
\> LRAs in dimension $h$, \\ 
$D_{h}=\frac{W_{h}}{\sum_{i\in \mathcal{L}}\left\vert R_{i}\right\vert }$ \> for the average demand of all \\
\> LRAs in dimension $h$,\\
$\lambda _{h}=\frac{W_{h}}{\sum_{k=1}^{d}W_{k}}$ \> for the normalized demand of \\
\> all LRAs in dimension $h$.
\end{tabbing}

\begin{table}
\centering
\caption{Application-Centric size measures $s_i$ for applications $i \in \mathcal{L}$}\label{tab:CombinedSizeMeasures}
\begin{tabular}{|lc|c|}
\toprule
Average & & $s_i = \frac{1}{d} \sum_{h=1}^{d}{s_{ih}}$ \\
\midrule
Max & & $s_i = \max_{1 \leq h \leq d}\{s_{ih}\}$ \\
\midrule
\makecell[l]{Average with\\exponential weight} & \cite{Panigrahy} &
$s_i = \sum_{h = 1}^{d}{e^{\varepsilon D_h} \cdot s_{ih}}$ \\
\midrule
Surrogate & \cite{CapraraToth-2001} & $s_i = \sum_{h = 1}^{d}{\lambda_h  s_{ih}}$ \\
\midrule
Extended Sum & \cite{LRASched2021} & $s_i =  \sum_{h=1}^{d}{ \frac{|R_i|}{W_h} s_{ih} }$ \\
\bottomrule
\end{tabular}
\end{table}

The \textit{Average measure with exponential weight} is one of the best performing measures in experiments on Vector Bin Packing, performed by Panigrahy et al.~\cite{Panigrahy}. It is computed as the weighted sum of $s_{ih}$-values, with exponential weights depending on average demands $D_h$. Parameter $\varepsilon$ is a small number selected appropriately for scaling. 

The \textit{Surrogate} measure is a natural extension of the 2-dimensional measure of Caprara and Toth~\cite{CapraraToth-2001}. It is computed as the weighted sum of $s_{ih}$-values, with the normalized demands $\lambda_h$ used for weights.

Finally, the \textit{Extended Sum} is an adaptation of the measure used in \textsc{Lra}Sched~\cite{LRASched2021}.
For application $i$, it is defined as the sum, over all dimensions $h$, of the demands of all replicas of that application $|R_i|s_{ih}$ in dimension $h$ normalized by the total demand $W_h$ of all applications in that dimension.

Prior research in the area of Bin Packing with Conflicts has discovered the benefits of combining the demand-based measure $s_i$ with the conflict-based measure, which takes into account the criticality of an application in terms of interference~\cite{Muritiba2010}.
Generalizing these ideas to affinity restrictions {\bf [R2]} of our model, we define the {\it hybrid demand-affinity} measure as the weighted sum of the demand-based measure $s_i$ and the affinity-based measure $\delta_i$: 
\begin{equation}
\Tilde{s}_{i}=\alpha \frac{s_{i}}{\overline{s}}+\left( 1-\alpha \right)
\frac{\delta _{i}}{\overline{\delta }}.   \label{hybrid-sizes}
\end{equation}
Here $s_i$ is computed via one of the expressions from Table~\ref{tab:CombinedSizeMeasures}, $\delta_i$ is the total number of applications linked with application $i$ in the affinity graph, while $\alpha \in[0,1]$ is chosen to give a higher priority to application demands or to interference.
Scaling is performed for handling incomparable parameters, dividing by $\overline{s}$ and $\overline{\delta}$, the average values of $s_{i}$ and $\delta_{i}$, respectively.\\

\myparagraph{Nodes' residual capacities} The key characteristics of a partly loaded node $n\in \mathcal{N}$ are residual capacities $\overline{C}_{nh}$, maintained for all dimensions $h=1,2, \ldots, d$.
They are computed as original node capacities $C_{h}$ minus the total size of allocated replicas for the same dimension $h$.
In the presence of residual capacities in multiple dimensions, 
there is a need to compute a single \textit{residual capacity measure} $\overline{C}_{n}$ for each node $n \in \mathcal{N}$ and to use it as a priority index.

For each application size measure $s_{i}$ from Table~\ref{tab:CombinedSizeMeasures}, we similarly define the corresponding node measure $\overline{C}_{n}$.
To this end, we replace in each formulae the size $s_{ih}$ of application $i$ in dimension $h$ by the residual capacity $\overline{C}_{nh}$ of the node $n$ in the same dimension, and adjust calculation of $W_h, D_h, \lambda_h$ accordingly.
In all calculations we replace $|R_i|$ by $1$. This way  $|R_i| s_{ih}$ representing the demand of application $i$ in dimension $h$ is replaced by $\overline{C}_{nh}$, the residual capacity of node $n$ in dimension $h$. Also, $\sum_{i \in \mathcal{L}} |R_i|$ representing the total number of all replicas is replaced by $|\mathcal{N}|$, the total number of activated nodes. \\

\subsection{Node-Centric Algorithms}\label{sec:solution:node-centric}

For the Node-Centric approach, the {\it application-node score} for application $i$ and node $n$, denoted by  $\xi_{in}$, is computed only for a {\it feasible} application node pair.
The higher the score, the more beneficial it is to allocate replicas of application $i$ to node $n$, which is currently being packed. 

\begin{table}
\centering
\caption{Bin-Centric scores $\xi_{in}$ for applications $i \in \mathcal{L}$ and nodes $n \in \mathcal{N}$}\label{tab:scores}
\begin{tabular}{|lc|c|}
\toprule
DotProduct & \cite{Panigrahy} & $\xi_{in} = \sum_{h=1}^{d} {s_{ih} \overline{C}_{nh}}$ \\
\midrule
L2Norm & \cite{Panigrahy} & $\xi_{in} = -\sum\nolimits_{h=1}^{d} {\left ( \overline{C}_{nh} - s_{ih} \right)^2}$ \\
\midrule
Fitness & \cite{LRASched2021} & $\xi_{in} = \sum_{h=1}^{d}{\frac{s_{ih}}{W_h} \cdot \frac{\overline{C}_{nh}}{\sum_{k\in\mathcal{N}}{\overline{C}_{kh}}}}$ \\
\midrule
TightFill & & $\xi_{in} = \sum_{h=1}^{d} {\frac{s_{ih}}{\overline{C}_{nh}}}$ \\
\bottomrule
\end{tabular}
\end{table}

We explore in our algorithms the known best-performing scores, together with a newly proposed score, denoted by \textit{TightFill}, as shown in Table~\ref{tab:scores}.

All four scores select for a current node $n$ the application which uses the $d$ resources of the node to the highest extent. 
\begin{itemize}[leftmargin=1.5em]
    \item In the {\it DotProduct} score this is achieved by prioritizing the dimensions for which node $n$ has the largest capacity. An application with highest demands in those dimensions is considered as the best choice.
    \item In the {\it L2Norm} score, the expression is negative so that the smallest positive value indicates the best application for node $n$.
    The preferred application minimizes the difference between its size and residual capacity of the node measured via the L2 norm.
    \item In the {\it Fitness} score, the application demands $s_{ih}$ are normalized with respect to $W_h$, the total demand of all applications in dimension $h$, and the node capacities $\overline{C}_{nh}$ are normalized with respect to the total free capacity of all nodes in dimension $h$, $1 \le h \le n$.
    \item The \textit{TightFill} score is a counterpart of \textit{DotProduct} which ensures the tightest usage of the node residual capacity.
\end{itemize}

\subsection{Multi-Node Algorithms} \label{ssec:MultiNode}
Recall that multi-node algorithms require a target number of nodes as part of the input.
The search for a feasible solution with the minimum number of nodes is arranged by calling the algorithm repeatedly with different trial values for the number of nodes, either via binary search or with the trial value decremented in steps.\\

\myparagraph{Multi-Node Algorithms with Replica Spreading}
These algorithms use the same principles as the Application-Centric algorithms, but with the aim of replica spreading across the whole pool of activated nodes, reducing this way the restrictions imposed by the affinity constraints $a_{ij}$.
Among the six Application-Centric algorithms discussed in Section~\ref{sec:solution:app-centric}, only Worst Fit and Worst Fit Decreasing produce different solutions if $n$ nodes are activated at start rather than being activated one by one on the fly.
Indeed, after allocating a replica to a node, that node is placed further down in the list, and in the next step another node is selected to allocate next replica, thus enabling replica spreading over multiple nodes. The remaining Application-Centric algorithms, First Fit, First Fit Decreasing, Best Fit and Best Fit Decreasing, do not change their behavior if a pool of nodes is activated at start. For this reason, we create only two algorithms by combining the Multi-Node and the Application-Centric approaches, with the shortcut names \textit{SpreadWF} and \textit{SpreadWFD}. \\

\myparagraph{Multi-Node Algorithms with Application-Node Matching}
These algorithms use the same principles as the Node-Centric algorithms, but on a pool of $n$ activated nodes rather than on single nodes considered one by one.
Each time, the most appropriate application-node pair is selected among all possible pairs of unallocated applications and non-fully packed nodes by using the scores defined in \S\ref{sec:solution:node-centric} for the Node-Centric approach, and a single replica is allocated.
It is expected that the replicas of an application are spread broadly across the nodes pool, with less restrictions caused by the affinity constraints.

\begin{table}[t]
\centering
\caption{Algorithms' time complexity. $L$ is the number of applications, $R$ is the total number of all replicas of all applications, $n$ is the given (target) number of nodes}
\label{tab:RunningTimes}
\begin{tabular}{|ll|}
\toprule
Application-Centric & $O(R^2 L)$\\
\midrule
Node-Centric & $O(R L^2)$ \\
\midrule
\makecell[l]{Multi-Node with Replica Spreading\\~~and $n$ nodes} & $O(RLn)$\\
\makecell[l]{Multi-Node with Application-Node\\~~Matching and $n$ nodes} & $O(R L^2 n)$\\
\bottomrule
\end{tabular}
\end{table}

\subsection{Time Complexity of Algorithms}
\label{ssec:Tuning}

The three introduced approaches, Application-Centric, Node-Centric and Multi-Node, provide the foundation to build a wide range of heuristics.
The choice of a specific method, together with the most appropriate measures or scores, depends on special features of scenarios and datasets, and on limitations on algorithms' running times.
Analytical estimates of running times are provided in Table~\ref{tab:RunningTimes}.
Clearly, for large-size datasets, the Multi-Node algorithms with Application-Node Matching may become unacceptably slow.
Note that the actual performance of the algorithms may differ from the theoretical estimates since the worst-case analysis takes into account very rare scenarios.
For example, in the analysis we assume that the number of activated nodes is $O(R)$, which is a highly pessimistic estimate.
Similarly, in the absence of more accurate information, we assume that there can be up to $O(R)$ replicas in a single node in the worst case.
Both assumptions affect the time complexity of the most frequent step of any algorithm, the feasibility check.
Note also that the running time estimates for the Multi-Node approach are made for a single call with a fixed $n$ given as the trial number of nodes.
These estimates have to be multiplied by the total number of calls made by the decrementing method, or by the binary search, to get the time complexity of the overall procedure.

The choice of the size measure (Table~\ref{tab:CombinedSizeMeasures}) for $s_i$ should take into account not only its impact on the running time, but also the nature of the dataset.
In the presence of a dominating (bottleneck) resource type $h^*$, $1 \le h^* \le d$, which plays the critical role in application allocation, computing of the measure $s_i$ can be simplified by using $s_i = s_{ih^*}$ instead. 
For Application-Centric approaches, incorporating the hybrid size measure of Eq.~(\ref{hybrid-sizes}) on top of one of the standard measures from Table~\ref{tab:CombinedSizeMeasures} can be beneficial if affinity constraints are very restrictive, so that many pairs of applications are in conflict.
Note that Eq.~(\ref{hybrid-sizes}) does not affect the asymptotic worst-case time complexity, but may slow down the algorithms' performance on large datasets.

\subsection{Classification of Existing Schedulers}\label{sec:solution:sota}

We conclude this section by outlining the published approaches used for scheduling LRAs with affinity restrictions, which will serve as baselines in the experiments: two heuristics of 
Medea~\cite{garefalakis2018medea} and one heuristic of \textsc{Lra}Sched~\cite{LRASched2021}.

The \textit{TagPopularity} (Medea-TP) heuristic is Application Centric. It allocates applications one by one, starting with those having the highest interference.
This heuristic can be classified as FFD with size measure $s_i=\delta_i$, the special case of Eq.~(\ref{hybrid-sizes}) with $\alpha = 0$.

\textit{NodeCandidates} (Medea-NC) is another version of the Application-Centric approach, with $s_i$-parameters representing the total number of available nodes in the system which can accommodate a replica of $i$, observing capacity and affinity restrictions:
\begin{equation}
s_i = \sum_{n \in \mathcal{N}} \zeta_{in}. \label{size-nodecount}
\end{equation}
Here $\zeta_{in} =1$ if a replica of application $i$ can be allocated to node $n$ without violating affinity restrictions, and $\zeta_{in} = 0$, otherwise.
Applications are allocated one by one, starting with the most restrictive ones, i.e., those having the lowest sizes $s_i$ computed by Eq.~(\ref{size-nodecount}), and sizes of the remaining applications are re-computed after each allocation step.

\textsc{Lra}Sched~\cite{LRASched2021} uses a two-phase approach. The first phase aims at maximizing the number of fully allocated LRAs and resource utilization of the given restricted pool of available nodes. The second phase aims at minimizing the number of new nodes used to allocate remaining LRAs.
The second phase employs a Node-Centric algorithm with the \textit{Fitness} score. 
We denote the algorithm of this second phase by \textit{LRASched-Fitness}.

\section{Performance Evaluation}
\label{sec:exp}

All algorithm codes, scripts for generating the instances, as well as additional figures, are publicly available at \texttt{\small
\url{https://github.com/DSSGroup-Leeds/LRA-binpacking-expe}}.

\subsection{Experimental Settings}
\label{sec:exp:setting}

\begin{table*}[t]
\centering
\caption{Summary of generated instances}\label{tab:NewInstances}
\begin{tabular}{|c|c|c|}
\toprule
Scenario & \makecell{Varied affinity density} & \makecell{Varied number of LRAs} \\
\midrule
$\left\vert \mathcal{L}\right\vert $ & $9,338$ & \makecell{$10,000$\\$50,000$\\$100,000$} \\
& & \\
$\left\vert \mathcal{R}_{i}\right\vert, s_{ih}$ & \makecell{same as Alibaba~\cite{tianchidata}} & \makecell{similar to Alibaba~\cite{tianchidata}} \\
& & \\
affinity density $\Delta$ & \makecell{$1\%$\\$5\%$\\$10\%$} & $0.5\%$ \\
& & \\
affinity graph type & \makecell{arbitrary\\threshold\\normal} & \makecell{arbitrary\\threshold\\normal}\\
\midrule
\midrule
\makecell{$d=2$\\(CPU, memory)} & \makecell{90 instances \\without temporal changes} & \makecell{90 instances \\ without temporal changes} \\
\midrule
\makecell{$d=98 \times 2$\\(CPU, memory, $98$ epochs)} & \makecell{90 instances \\ with temporal changes} & \makecell{90 instances \\ with temporal changes} \\
\bottomrule
\end{tabular}
\end{table*}

\myparagraph{Simulation configuration and instance generation}
As the pre-execution planning is independent from the runtime execution of LRAs, we adopt simulation-based evaluation to validate the efficacy of different algorithms on a single machine equipped with one Intel Xeon Gold 6138 CPU and 64 GB of memory. We simulate different scales of LRA submission and evaluate how our algorithms succeed in LRA allocation onto a mocked compute cluster with identical nodes comprising 64 CPU cores and 128 GB of memory.

Our aim is to examine several sets of instances, each set with common features and related to a specific scenario, and to select the winning algorithms from our suite.
The instances stem from the dataset published by the Alibaba Tianchi Platform~\cite{tianchidata}.
The original dataset contains the data for $9,338$ LRAs with a total of $68,224$ replicas and $24,078$ affinity restrictions. Each LRA has resource requests in two dimensions: CPU cores and memory. LRA resource profiles change over time, with recordings known for 98 time sampling points.

We study two scenarios: one with different densities of affinity restrictions and another one with different numbers of LRAs. Our aim is to evaluate the impact of these characteristics on the solution quality and the running times of the proposed algorithms. Each scenario is subdivided into two sets of instances depending on whether LRA resource requests are constant or change over time. Each set contains a total of $90$ instances:
\begin{itemize}[leftmargin=1.5em,noitemsep]
    \item three types of affinity graphs (arbitrary, normal, threshold),
    \item three values of one of the varied parameters (affinity density or the number of LRAs),
    \item 10 instances for each combination.
\end{itemize}
A summary of the generated instances is presented in Table~\ref{tab:NewInstances}.

In the instances with \textit{varied affinity density}, represented in the second column of Table~\ref{tab:NewInstances}, the number of LRAs $| \mathcal{L} |$ is the same as in the original Alibaba dataset~\cite{tianchidata}, while the number of affinity restrictions, measured as \textit{affinity density}, is different.
The affinity density $\Delta$ is defined as the average number of affinity restrictions per LRA divided by the total number of LRAs.
For example, affinity density of $10\%$ means that each LRA has affinity restrictions with $10\%$ of other LRAs on average.
Note that in the original Alibaba dataset, the affinity density is lower than $0.05\%$.
We select higher density values for experiments to investigate the impact of affinity restrictions on the solution quality and algorithms' running times. 
For each LRA, the number of replicas per application $| \mathcal{R}_i |$ and resource requirements $s_{ih}$ are kept unchanged, as in the original Alibaba dataset.

In the instances with \textit{varied number of LRAs}, represented in the third column of Table~\ref{tab:NewInstances}, the affinity density is fixed to the same value ($0.5\%$), while the number of LRAs $| \mathcal{L} |$ is different.
We select larger instances compared to the Alibaba dataset~\cite{tianchidata} to explore the capabilities of the algorithms for optimizing the performance of massive scale systems.
The values for the number of replicas $| \mathcal{R}_i |$ and resource requirements $s_{ih}$ are defined using the same probability distributions as in the original Alibaba dataset.  

For any type of instance, affinity values $a_{ij}$ were generated following the same probability distribution as in the original Alibaba dataset.\\

Consider now the three approaches to graph generation, given number $|\mathcal{L}|$ of vertices and expected density $\Delta$. The method for generating \textit{arbitrary} graphs is described by Sadykov and Vanderbeck~\cite{Sadykov2013}. The method for generating \textit{threshold} graphs is described by Gendreau et al.~\cite{Gendreau2004} and elaborated further by Bacci and Nicoloso~\cite{ConflictErrata2017} for parameter correction. We propose the third approach to generate so called \textit{normal} graphs. It starts with a graph of $|\mathcal{L}|$ vertices and no arcs, and then for each vertex $i$ it randomly picks a value $p_i$ following the normal distribution of mean $\Delta |\mathcal{L}|$ and standard deviation $\Delta |\mathcal{L}| / 2$, restricting the value between $0$ and $|\mathcal{L}|-1$. Then $p_i$ vertices are selected at random using uniform distribution and they are used as end-nodes for arcs originating from vertex $i$. 

The resource requirements of each LRA are copied from the Alibaba dataset for all 98 sampling points, if considering  the class \textit{with temporal changes} (last row of Table~\ref{tab:NewInstances}), or they are extrapolated if considering the class \textit{without temporal changes} (penultimate row of Table~\ref{tab:NewInstances}): for each LRA $i$ we select the maximum values $s_{i1}$, $s_{i2}$ among those provided for the 98 sampling points and round them to the next integer. \\

\myparagraph{Evaluation methodology and metrics}
We evaluate the effectiveness and time efficiency of each algorithm. 

The \textit{effectiveness} is measured by recording  the number of nodes found in a feasible solution and calculating the \textit{deviation from the lower bound}, a ``lower-the-better'' indicator.
Since the total number of nodes cannot be smaller than the total demand $W_h$ of all LRAs in dimension $h$ divided by the node capacity $C_h$ in that dimension, where $h=1, \ldots,n$, the lower bound is defined as 
\begin{equation}
LB = \max_{1 \leq h \leq d} \left \{ \left \lceil \frac{W_h}{C_h} \right \rceil \right \}.\label{eq:LB}
\end{equation}

The \textit{time efficiency} is measured as the algorithm's computation time, averaged over the 10 instances of a given configuration of graph class and density value, or graph class and LRA number.\\

\myparagraph{Algorithm naming}
We implemented our algorithms and the three baseline algorithms, \textit{Medea-TP}, \textit{Medea-NC} and \textit{LRASched-Fitness} (\S\ref{sec:solution:sota}), in C++.

The shortcut names of \underline{Application-Centric} algorithms include the ordering rule (\S\ref{sec:solution:app-centric}) and the size measure (Table~\ref{tab:CombinedSizeMeasures}). For example, \textit{WFD-AvgExp} denotes the WFD algorithm with the size measure ``average with exponential weight''. 

\underline{Node-Centric Algorithms with Decreasing Scores} are denoted by \textit{NCD} followed by the scoring name (Table~\ref{tab:scores}). ``Decreasing score'' indicates the choice of the largest application-node score in each step. For example, \textit{NCD-DotProduct} denotes the Node-Centric algorithm with decreasing dot-product score.

Considering \underline{Multi-Node algorithms}, we focus on the versions with \textit{replica spreading} and exclude the versions with \textit{application-node matching} from our experiments, as their running times were observably too long even for the instances with 9,338 LRAs.

For the replica spreading versions we use prefix \textit{Spread} in the notation, and postfix \textit{BinSearch} or \textit{Decr}, depending on the search strategy used for multiple calls with different values of the target number of nodes. 

\textit{Binary search} strategy narrows down the interval which estimates the minimum number of nodes. It uses Eq.~(\ref{eq:LB}) for the initial lower bound, and the output of the \textit{First Fit (FF)} algorithm for the initial upper bound. For example, \textit{SpreadWFD-Avg-BinSearch} denotes the spreading version of WFD (with ``average'' size measure) in combination with binary search. 

The alternative, \textit{Decrementing} approach arranges the search by decreasing the target number of nodes in steps. For the starting point, it uses the same value for the upper bound as binary search. In the notation, postfix \textit{Decr} is followed by the step value. For example, \textit{SpreadWFD-Avg-Decr2} denotes the spreading version of WFD (with ``average'' size measure) in combination with the decrementing approach, which decreases the target number of nodes from the best value found so far, in decrements computed as $2\%$ of the lower bound.

\subsection{Capabilities of the ILP Model} \label{sec:exp:ILP}
The instances introduced in Table~\ref{tab:NewInstances} appeared to be too hard for the ILP model formulated in \S\ref{sec:model:form}. Considering smaller instances, we have found that solutions can be obtained for medium size instances, with up to $2,000$ LRAs having about $15,500$ replicas in total. In those instances, LRAs have resource requirements in CPU and memory, which do not change over time. This is the two-dimensional case of the problem under study. 
Allowing sufficiently large computation time, of up to 4 hours, Gurobi solver can find solutions within $0.2\%$ from lower bounds.

Clearly, for instances with more than $2,000$ LRAs, heuristics should be preferred due to their scalability and flexibility of integrating with real-life schedulers. 

\subsection{Results for Instances without Temporal Changes} \label{sec:exp:eff1}

\begin{figure}[t]
    \centering
    \includegraphics[width = 0.49\textwidth]{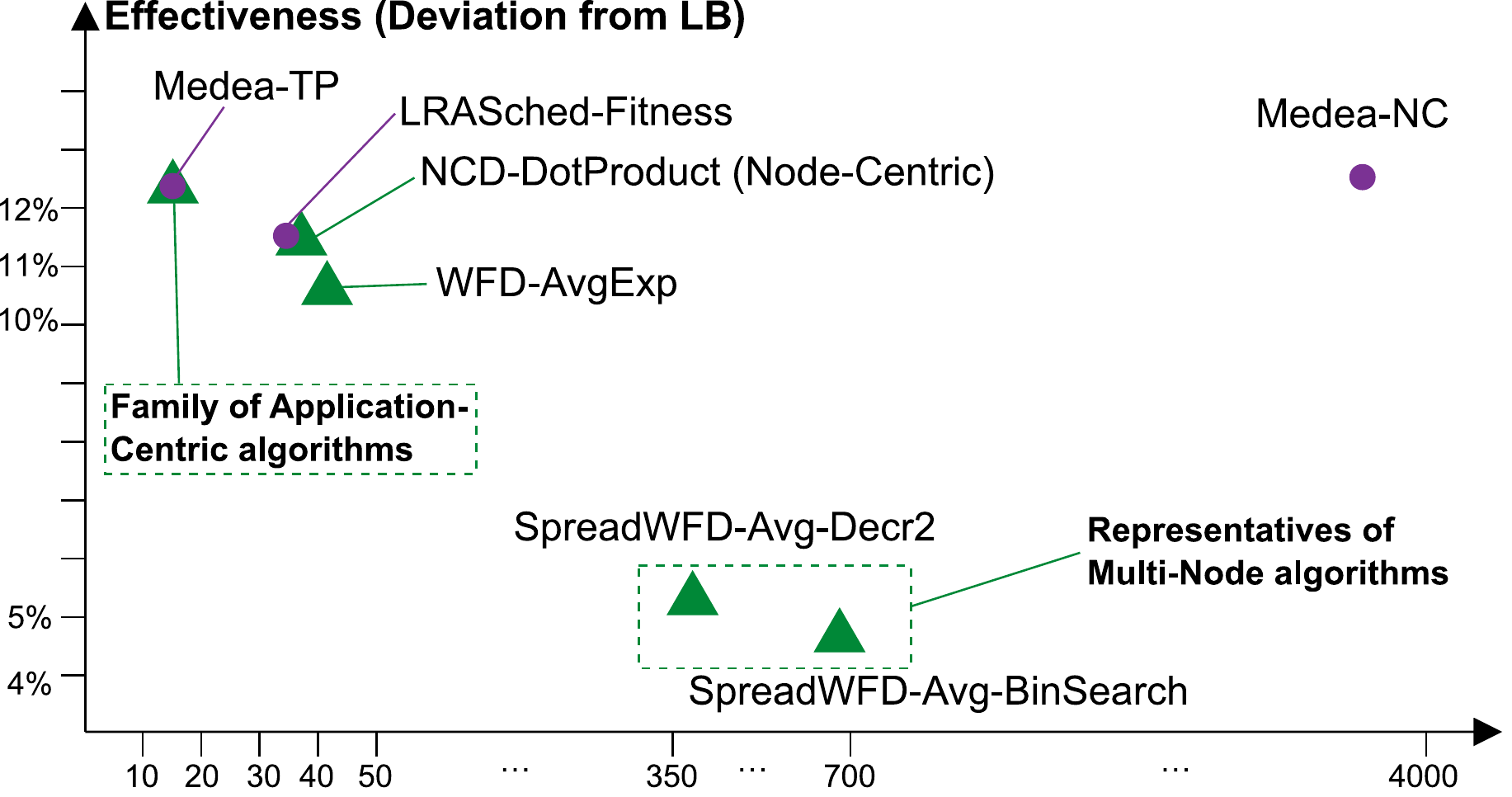}
    \vspace{-0.4em}
    \caption{Performance summary of algorithms for instances with $9,338$ LRAs, different affinity densities and without temporal changes}
    \label{fig:summary}
\end{figure}

In this section we discuss the performance of the algorithms on two-dimensional instances, which correspond to the penultimate row of Table~\ref{tab:NewInstances}. A high-level overview of the results, averaged over all 90 instances with different affinity densities, is illustrated in Fig.~\ref{fig:summary}. The trade-off between effectiveness and computation time can help practitioners in selecting the algorithm that best fits their requirements.

In the following, we analyze in depth the algorithms' performance on instances with varied affinity density (described in column 2 of Table~\ref{tab:NewInstances}) and on instances with varied number of LRAs (described in column 3 of Table~\ref{tab:NewInstances}). As no major differences were observed between the results obtained for the three types of affinity graphs, we report the results for the graphs of \textit{arbitrary} type, unless specified. \\

\begin{figure*}[t]
\centering
\subfigure[Effectiveness]{
\includegraphics[width=0.8\textwidth]{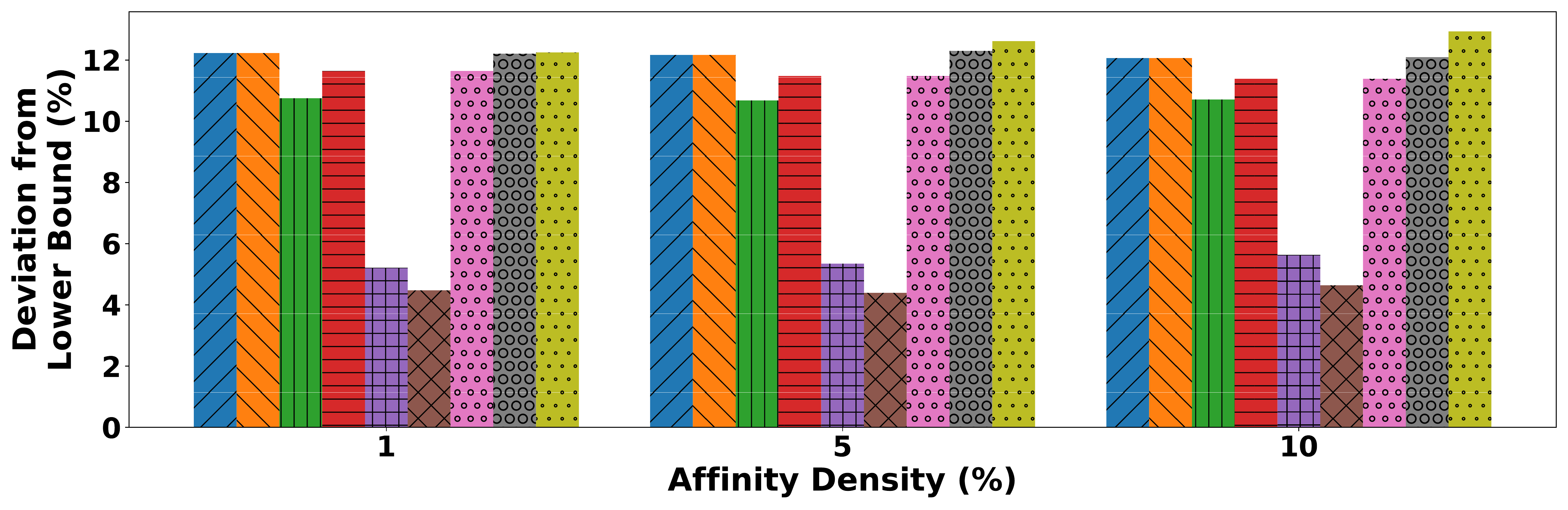}
\label{fig:density2D-result}
}
\hspace{-1em}
\subfigure[Time consumption - log-scale]{
\includegraphics[width=0.8\textwidth]{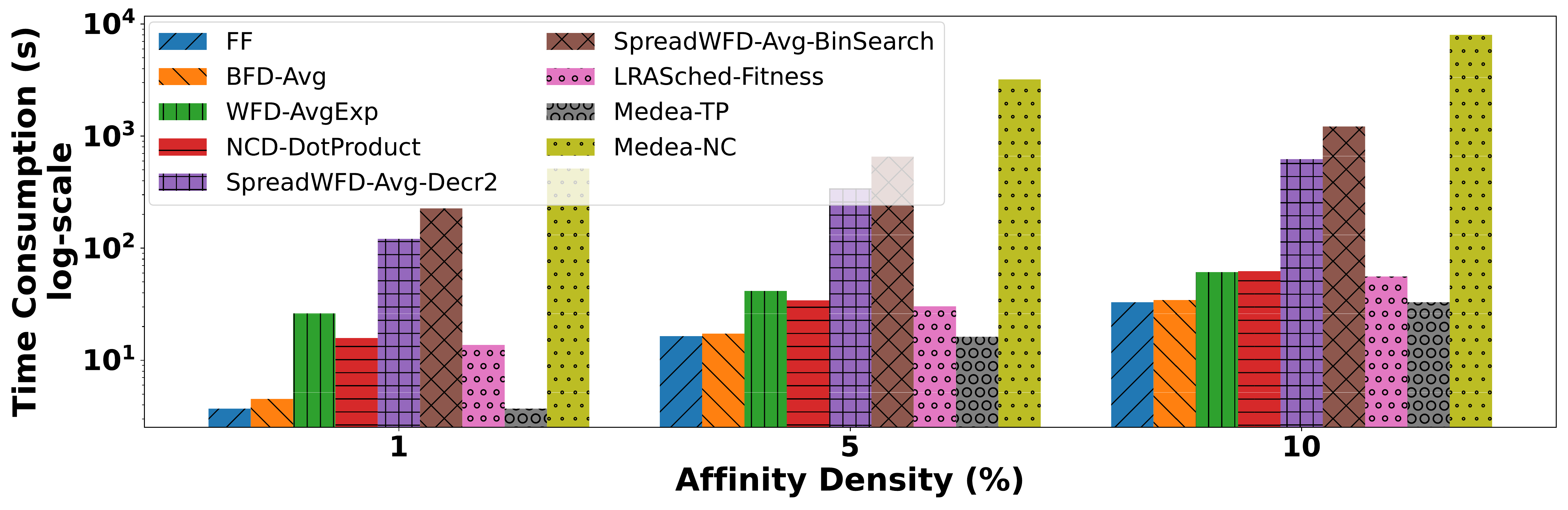}
\label{fig:density2D-time}
}
\vspace{-1em}
\caption{Different affinity densities under fixed resource requests, $|\mathcal{L}| = 9,338$}
\vspace{-0.4em}
\label{fig:density2D}
\end{figure*}

\begin{figure*}[t]
\centering
\subfigure[Effectiveness]{
\vspace{-1.4em}
\includegraphics[width=0.8\textwidth]{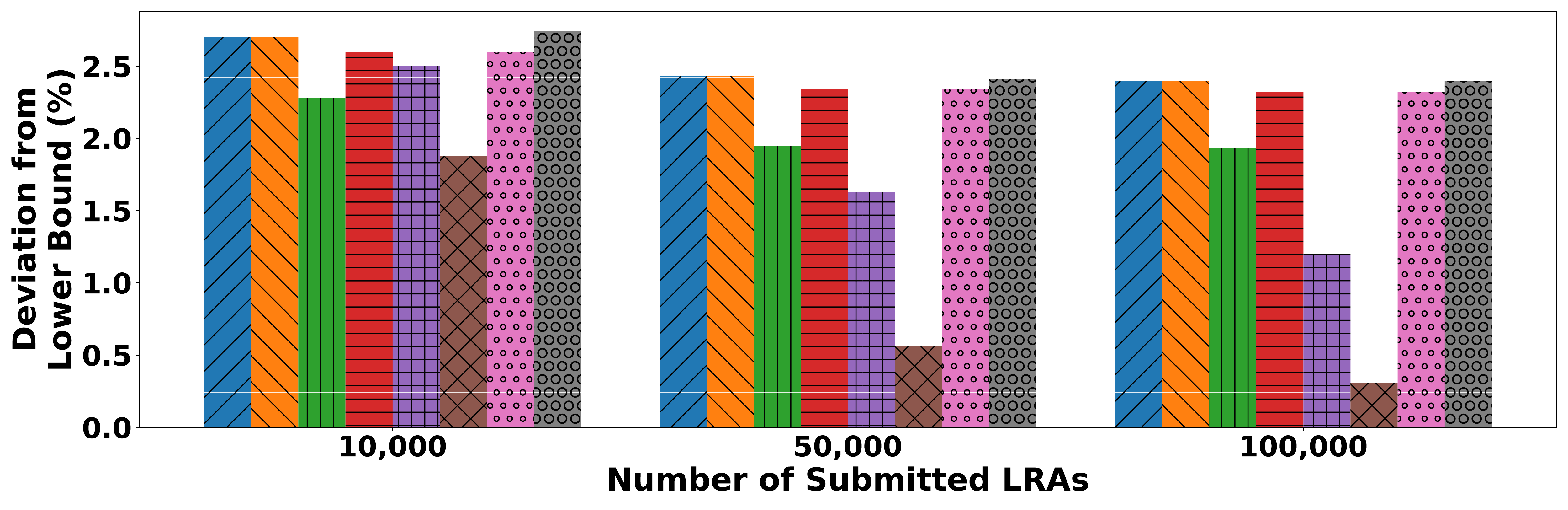}
\label{fig:large2D-result}
}
\hspace{-0.8em}
\subfigure[Time consumption - log-scale]{
\vspace{-1.4em}
\includegraphics[width=0.8\textwidth]{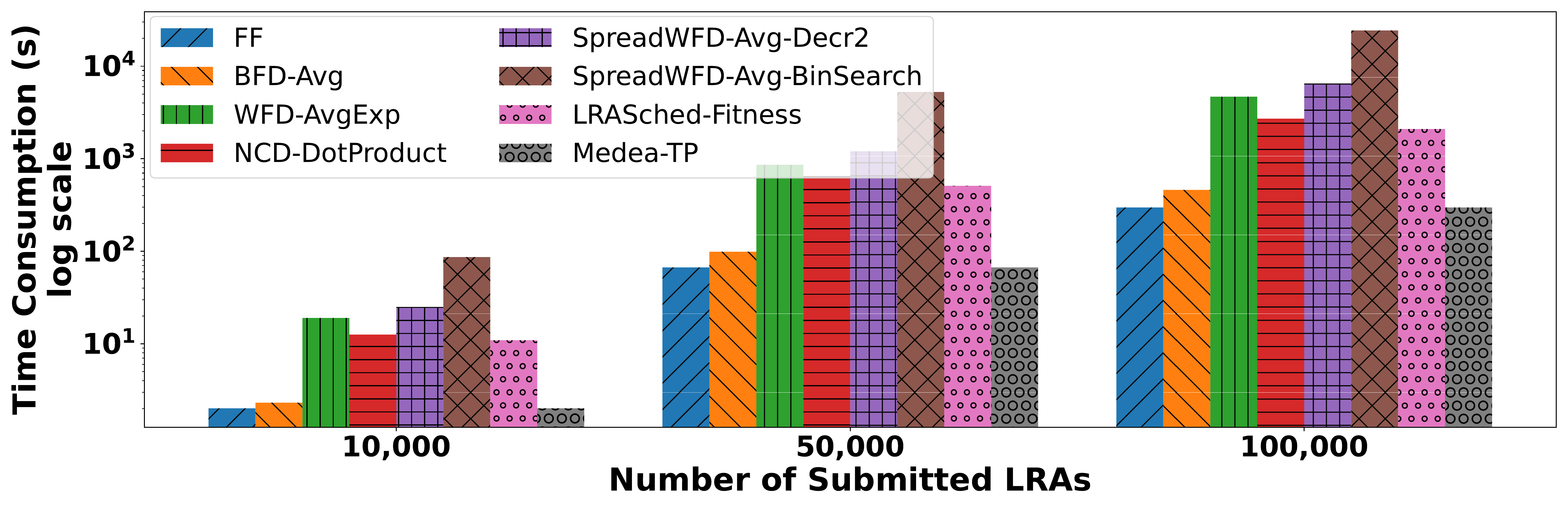}
\label{fig:large2D-time}
}
\vspace{-1.15em}
\caption{Different LRA numbers under fixed resource requests, affinity graph density is $0.5\%$}
\vspace{-0.4em}
\label{fig:large2D}
\end{figure*}

\myparagraph{Effectiveness}
\underline{\it Instances with varied density.}
In general, all Application-Centric algorithms (FF and various versions of FFD, BFD and WFD with different size measures) perform similarly, with approximately $12.1\%$ deviation from the lower bound on average, with two exceptions.
First, algorithms FFD, BFD and WFD with the ``Extended Sum'' measure are consistently worst-performing, with $15.6\%$ deviation on average.
Second, \textit{WFD-AvgExp} has $10.7\%$ deviation on average and consistently outperforms all others.
The advantage of \textit{WFD-AvgExp} stems from the focus on the most demanding dimensions when selecting the next LRA to be allocated. 

Node-Centric algorithms place themselves between \textit{WFD-AvgExp} and the other Application-Centric algorithms, with $11.5\%$ deviation on average.

The spreading versions of the Multi-Node algorithms are particularly successful. For example,  \textit{SpreadWFD-Avg-BinSearch} and \textit{SpreadWFD-Avg-Decr2} achieve $4.5\%$ and $5.4\%$ deviation from the lower bound, respectively. Solutions of similar quality are obtained by the versions of \textit{SpreadWFD-AvgExp}, but at the cost of larger computation time (a consequence of computing a more elaborate size measure).

We visualize the results of the representatives of each algorithm family in Fig~\ref{fig:density2D}, where we also include the summary of the baseline algorithms. We observe that \textit{Medea-NC}, with $12.6\%$ deviation, is outperformed by all other algorithms (except for algorithms with the ``ExtendedSum'' measure not included in Fig~\ref{fig:density2D}), while \textit{Medea-TP} performs similar to the Application-Centric algorithms, with $12.2\%$ deviation. \textit{LRASched-Fitness} works similar to other Node-Centric algorithms, with a slightly smaller execution time.
Compared with these baselines, our algorithms of type \textit{SpreadWFD-Avg} are $7\%$ closer to the lower bound.
This marginal number implies about $350$ nodes saving, which is of significance for cost-effective and energy-efficient datacenters.

Comparing the results for different affinity densities we do not observe noticeable differences in the algorithms' effectiveness. The exceptions are \textit{SpreadWFD-Avg-BinSearch} and \textit{SpreadWFD-Avg-Decr2} applied to the instances with threshold graphs, where the deviation from the lower bound increases from $3.6\%$ to $10.5\%$ as the graph density increases.

\noindent\underline{\it Instances with varied LRA number.}
As shown in Fig.~\ref{fig:large2D-result}, the algorithms' effectiveness generally improves when the LRA scale increases.
With $100,000$ LRAs, \textit{FF}, \textit{BFD-Avg} and \textit{Medea-TP} achieve $2.5\%$ deviation from the lower bound on average, \textit{NCD-DotProduct} and \textit{LRASched-Fitness} achieve $2.4\%$ deviation, and \textit{WFD-AvgExp} reaches $2\%$ deviation.

\textit{SpreadWFD-Avg-BinSearch} and \textit{SpreadWFD-Avg-Decr2} are particularly successful, achieving $0.9\%$ and $1.8\%$ deviation on average, with figures as low as $0.3\%$  for \textit{SpreadWFD-Avg-BinSearch} when applied to instances with $100,000$ LRAs. 
However, an interesting anomaly was observed for  smaller instances, with $10,000$ LRAs: there were several instances with \textit{arbitrary} and \textit{normal} affinity graphs for which two \textit{SpreadWFD} algorithms could not find better solutions than \textit{FF}. Still the  performance of \textit{SpreadWFD} is the best even on small instances, if averaging the results of multiple experiments.\\

\myparagraph{Execution time}
\underline{\it Instances with varied density.}
Fig.~\ref{fig:density2D-time} shows the average execution times of the algorithms when applied to the instances with different affinity densities.
FFD-based algorithms are among the fastest, along with \textit{FF} and \textit{Medea-TP}, while BFD-based algorithms are slightly slower.
All these algorithms merely take less than $5$s, $18$s, and $33$s for densities $1\%$, $5\%$ and $10\%$, respectively.
In contrast, WFD-based algorithms are much slower, taking $26$s, $41$s and $61$s, respectively.

Node-Centric algorithms and \textit{LRASched-Fitness} are in-between: \textit{NCD-DotProduct} takes $16$s, $34$s and $62$s on average for the three densities, while \textit{LRASched-Fitness} runs a few seconds faster.

Overall, the relative difference in running times between these algorithms tends to decrease when the affinity density increases.
With $10\%$ density, the running times for the WFD-based algorithms are similar to \textit{LRASched-Fitness} and \textit{NCD-DotProduct}.

\begin{figure*}[t]
\centering
\subfigure[Effectiveness]{
\vspace{-1.4em}
\includegraphics[width=0.8\textwidth]{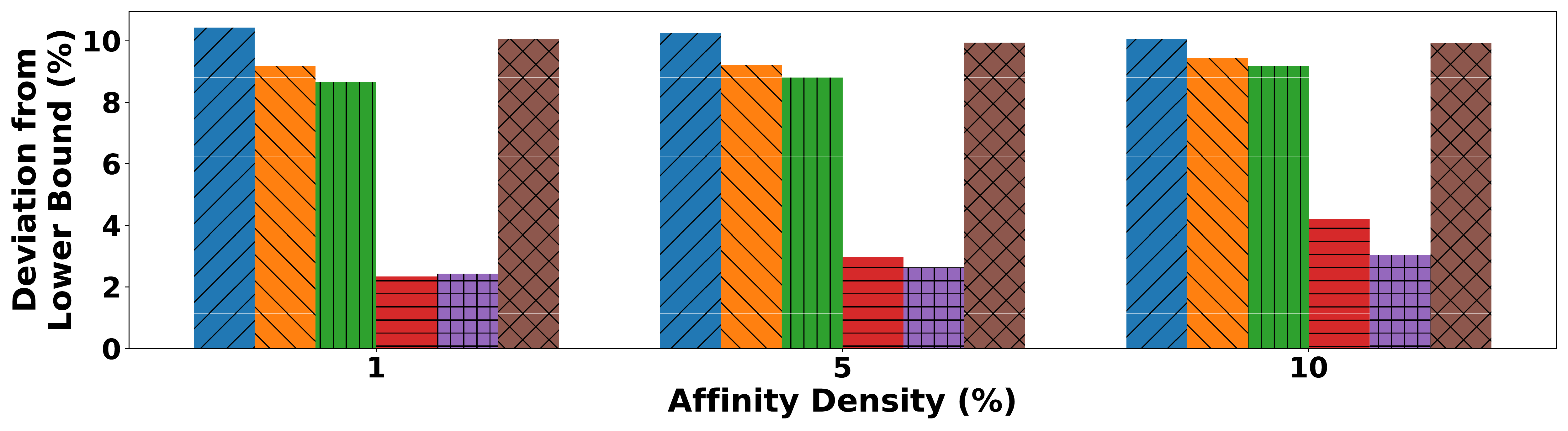}
\label{fig:densityTS-result}
}
\subfigure[Time consumption - log-scale]{
\vspace{-1.4em}
\includegraphics[width=0.8\textwidth]{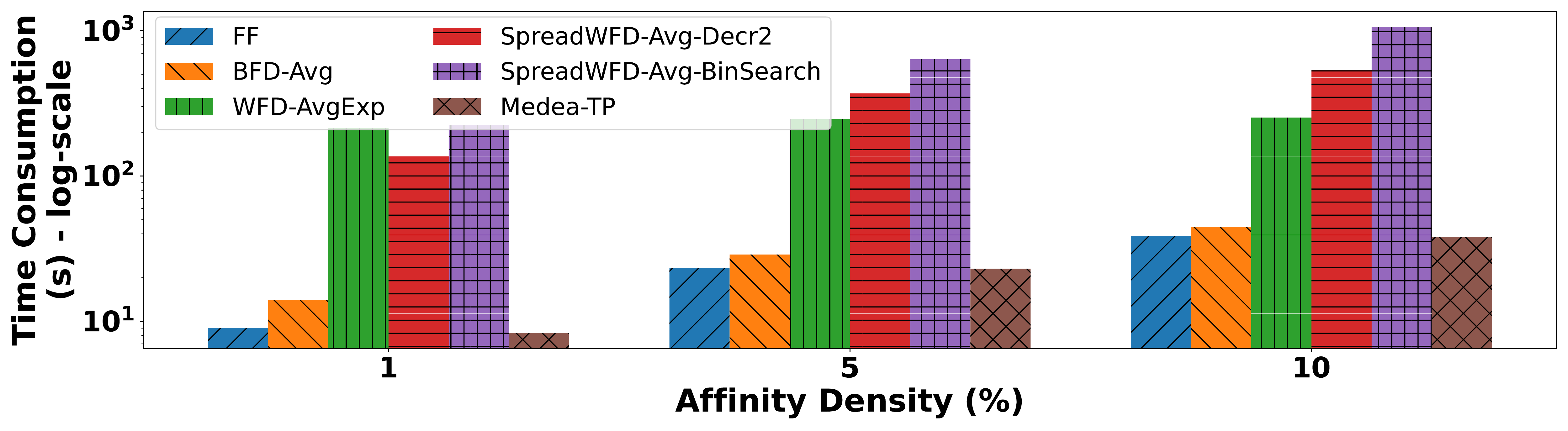}
\label{fig:densityTS-time}
}
\vspace{-1.6em}
\caption{Different affinity densities under time-varying resource requests, $|\mathcal{L}| = 9,338$}
\vspace{-0.4em}
\label{fig:densityTS}
\end{figure*}

\begin{figure*}[t]
\centering
\subfigure[Effectiveness]{
\vspace{-1.4em}
\includegraphics[width=0.8\textwidth]{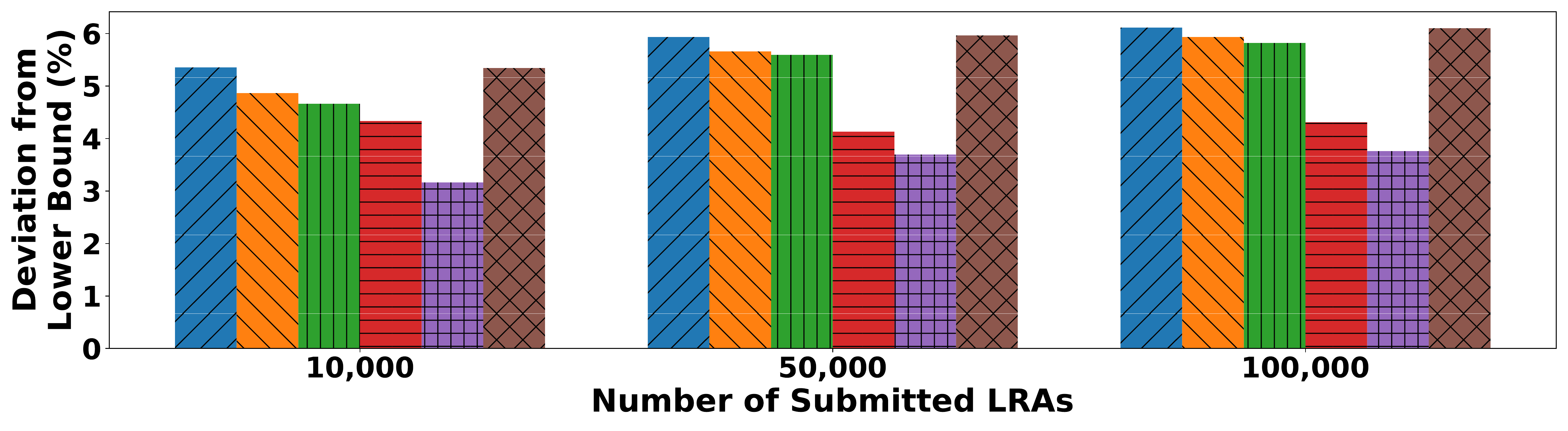}
\label{fig:largeTS-result}
}
\subfigure[Time consumption - log-scale]{
\vspace{-1.4em}
\includegraphics[width=0.8\textwidth]{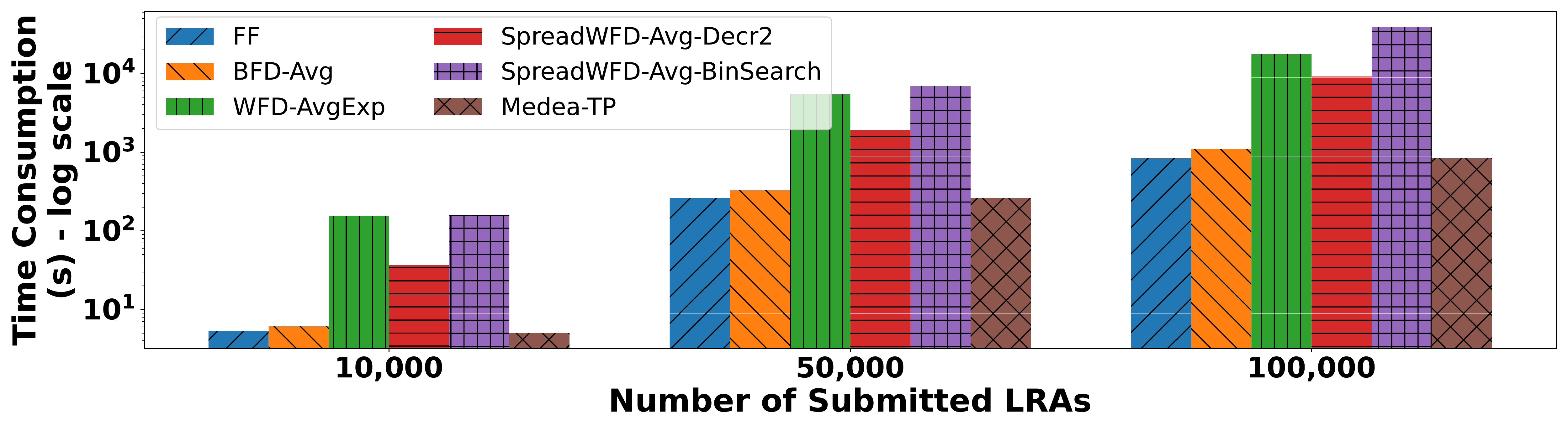}
\label{fig:largeTS-time}
}
\vspace{-1.6em}
\caption{Different LRA numbers under time-varying resource requests, affinity graph density is $0.5\%$}
\vspace{-0.4em}
\label{fig:largeTS}
\end{figure*}

The best-performing algorithm \textit{SpreadWFD-Avg-BinSearch} is unsurprisingly among the slowest algorithms, taking $225$s, $653$s and $1214$s on average, when the affinity density grows.
This is because binary search needs iterative calls of the replica spreading version of \textit{WFD} to find the appropriate number of nodes. 
Replacing binary search by iteratively decreasing the number of target nodes enables \textit{SpreadWFD-Avg-Decr2} to achieve a two-fold speedup, compared with the binary search version.

\textit{Medea-NC} is the slowest algorithm observed.
It takes on average $512$s, $3,200$s and $8,005$s when handling the instances with $1\%$, $5\%$ and $10\%$ density.
It is worth noticing that, while using fine-tuned data structure may reduce the running time of \textit{Medea-NC}, its effectiveness would not change and remain inferior to other algorithms.

\noindent\underline{\it Instances with varied LRA number.}
Fig.~\ref{fig:large2D-time} shows the average execution times of the algorithms applied to the instances with different numbers of LRAs, and obviously there is an increasing trend when there are more LRAs to be scheduled.
The fastest algorithms include \textit{FF}, \textit{Medea-TP} and \textit{BFD-Avg} that can solve instances with $100,000$ LRAs within $8$ minutes.
In contrast, \textit{LRASched-Fitness}, \textit{NCD-DotProduct} and \textit{WFD-AvgExp} are much slower, taking about $35$, $45$ and $78$ minutes on average to do the same task.
Spreading approaches take even longer time: $2$ and $7$ hours, respectively.
\textit{Medea-NC} was excluded from this series of experiments due to overly excessive execution time even for $10,000$ LRAs.
Aligned with Fig.~\ref{fig:summary}, the results indicate that datacenter operators need to thoroughly strike a balance between the targeted solution quality and the permitted planning time to pinpoint the bespoke option.

\subsection{Results for Instances with Temporal Changes}
The instances with time-varying resource requests of applications are modeled as the problem with $d=98 \times 2$ dimensions, as described in the last row of Table~\ref{tab:NewInstances}. 
This dimension increase leads to a substantial growth of execution time. \textit{Medea-NC}, \textit{LRASched-Fitness} and Node-Centric algorithms such as \textit{NCD-DotProduct} were discarded from the performance comparison for being too computationally expensive.
Again, as no major differences were observed between results of the three different affinity graphs, we only report the results for the graphs of \textit{arbitrary} type, unless specified.
\\

\myparagraph{Effectiveness}
\underline{\it Instances with varied density.}
For the majority of the algorithms, the change in the affinity density does not significantly affect the accuracy of the solutions found, as demonstrated in Fig~\ref{fig:densityTS-result}.
The exceptions occur for the threshold graphs, similar to the instances without temporal changes: there is a substantial degradation in the performance of the two \textit{SpreadWFD} algorithms, from $2.2\%$ to $10.1\%$ when the affinity density changes from $1\%$ to $10\%$.
Again, this is because the \textit{SpreadWFD} algorithms could not find better solutions than the given upper bound on several instances with $5\%$ or $10\%$ density, and the solutions from \textit{FF} were used instead.

\noindent\underline{\it Instances with varied LRA number.}
As shown in Fig.~\ref{fig:largeTS-result}, 
there is a negligible discrepancy among the performance of each algorithm with different numbers of LRAs, when handling time-varying resource requests.
For example, with \textit{SpreadWFD-Avg-BinSearch}, the deviation from the lower bound only increases from $3.2\%$ to $3.8\%$ when the LRA number grows from $10,000$ to $100,000$.
Similar observations are valid for other algorithms, indicating that the proposed algorithms are successful in large-scale scenarios.\\

\myparagraph{Execution time}
\underline{\it Instances with varied density.}
As shown in Fig.~\ref{fig:densityTS-time},
\textit{FF}, \textit{BFD-Avg} and \textit{Medea-TP} can solve any instance within $45$ seconds on average, while \textit{WFD-AvgExp} finishes within $4$ minutes and \textit{SpreadWFD-Avg-Decr2} within $9$ minutes.
\textit{SpreadWFD-Avg-BinSearch} takes about $18$ minutes to solve high density instances with $9,338$ LRAs, which seems to be the best choice of algorithm considering its achieved effectiveness of less than $3\%$ deviation from the lower bound, on average.
It is also worth noticing that, for instances with $1\%$ density, the running times of \textit{SpreadWFD-Avg-BinSearch} and \textit{WFD-AvgExp} are similar and almost double the running time of \textit{SpreadWFD-Avg-Decr2}. This is particularly unexpected for \textit{WFD-AvgExp}, which involves one call of the application-centric WFD-algorithm, compared to multiple calls of \textit{SpreadWFD-Avg-Decr2}. 

\noindent\underline{\it Instances with varied LRA number.}
As shown in Fig.~\ref{fig:largeTS-time}, similar but smaller differences in the execution times can be observed under different submission scales, compared with the observations in Fig.~\ref{fig:densityTS-time}. 
The disparity is due to the computation time of the size measures of LRAs with $196$ dimensions.
Numerically, \textit{FF} and \textit{Medea-TP} can solve any instance with $100,000$ LRAs in $14$ minutes on average and \textit{BFD-Avg} takes $18$ minutes. \textit{SpreadWFD-Avg-Decr2}, \textit{WFD-AvgExp} and \textit{SpreadWFD-Avg-BinSearch} take $2.5$, $5$ and $11$ hours, respectively, to solve the largest instances.
Interestingly, \textit{SpreadWFD-Avg-Decr2} appears to be the best choice for instances with time-varying resource requests, as it achieves effectiveness close to the best algorithm, \textit{SpreadWFD-Avg-BinSearch}, with a 4-fold speedup in terms of the running time.

\section{Algorithm Recommendations}  
\label{sec:choice}

We recommend Application-Centric algorithms if the computation time is required to be as small as possible. In that group of algorithms, the  version of the traditional bin packing algorithm \textit{First Fit} (FF), adjusted to handle the problem with replicas and affinities, is among the fastest approaches. Its solution quality is either similar or just slightly worse than the quality of solutions found by other Application-Centric algorithms. Only one published algorithm, \textit{Medea-TP}, achieves comparable computation time and solution quality. As we show in \S\ref{sec:solution}, \textit{Medea-TP} belongs to the same group of Application-Centric algorithms and differs from \textit{FF} by an additional ordering of LRAs. It appears that, on the instances generated from the Alibaba Tianchi dataset, special ordering does not have a significant impact on the quality of the solution and on computation time. 

We recommend Multi-Node algorithms if the primary aim is to find solutions of the best quality, possibly with longer but still acceptable computation time (say, up to 30 minutes to allocate $10,000$ LRAs). An ultimate winner in our experiments is \textit{Spread-WFD-Avg-BinSearch}. It uses a special spreading mechanism to allocate replicas of the same LRA across different nodes. The spreading mechanism substantially increases the range of nodes suitable for co-location of a current application with a broader set of compatible LRAs. Additionally, it adopts binary search to identify the smallest, but feasible, number of nodes in the solution. None of the algorithms, either in our  suite or among the published ones, achieves the same solution quality, namely $0.3\%$ deviation from the lower bound, when handling instances with $100,000$ LRAs. 

Finally, in-between the two extremes of fastest but less accurate algorithms, and slowest but most accurate ones, there are those of intermediate running time and intermediate solution quality. All Node-Centric algorithms fall into this category, with \textit{LRASched-Fitness} and \textit{NCD-DotProduct} being best performing. Both algorithms produce solutions of comparable quality and differ slightly in their running times: \textit{LRASched-Fitness} is faster on instances with affinities, while \textit{DotProduct} is faster and superior in terms of the solution quality on instances without affinities.

There is one outlier in the Application-Centric group, \textit{WFD-AvgExp}: it performs slower than the majority of Application-Centric algorithms and slower than the Node-Centric algorithms but outperforms all of them in terms of the solution quality. We would like to observe that overall the Application-Centric algorithm WFD is often overlooked by practitioners and not included in their trials.

As a final note, we observe that all algorithms become much slower for instances with time-varying profiles, and the Node-Centric algorithms become prohibitively slow. Therefore, we narrow down our recommendations to \textit{Medea-TP} and \textit{FF} (the fastest), \textit{Spread-WFD-Avg-Decr2} (of intermediate running time and solution quality) and \textit{Spread-WFD-Avg-BinSearch} (the best solution quality).

\section{Practical Considerations}

\myparagraph{Integration into multi-stage cluster management} 
While this paper focuses on the algorithmic support for resource provisioning, the proposed algorithm suite can be more widely integrated in a multi-stage cluster management that consists of cluster initialization and runtime scheduling. 

At the initialization stage, given that the scheduling system foreknows all information of LRAs to be submitted, the resource planner that runs the algorithm suite can work out the best option for scheduling the LRAs with the minimal required nodes.
Horizontal scaling will be consequently used to match the planning outcome, through elastically sizing the number of bare metal servers or virtual machines in the resource pool. 
Once the cluster is initialized for hosting the LRAs, the cluster management will shift into the runtime scheduling stage that responds to the new LRA submissions and available resource release.
Cluster schedulers can accept any incoming LRA in the regular round of resource allocation~\cite{verma2015large,yang2020performance,lo2015heracles,hu2020toposch}.
Consequently, the admitted LRAs will gradually consolidate the nodes in the cluster until there is no room for new LRAs and a long waiting queue manifests.
Cluster auto-scaling will be performed to mitigate the long starvation of the waiting LRAs and handle dynamic load spikes.
The resource provisioning algorithm will be re-triggered accordingly. 
 
\myparagraph{Runtime management considerations} While our algorithm suite can provide competitive solutions that minimize the number of required computing nodes, the resource provisioning in practice usually comes with some resource slack or over-provisioning to increase reliability for the unknown and prevent degradation in user experience. Based upon the calculation of initial resource provisioning as a guidance, additional resource reservation by system operators allows to mitigate uncertainties at runtime such as an excessive increase in LRA's tail latency, out-of-memory problems when the LRA's resource usage fluctuates, failures or stragglers due to unexpected data stream coming into the LRA, etc. 
The reserved yet idle resources can be harvested by using a series of system optimization techniques including hypervisor or kernel level oversubscription~\cite{yang2020performance,lo2015heracles,kim2014group} and core reassignment mechanism~\cite{ousterhout2019shenango}.

\myparagraph{Other objectives considerations}
The scheduling problem formulated in the paper is an attempt to find the minimum number of nodes that accommodate different LRA scales and affinity restriction densities.
However, in real scenarios, the compute capability is sometimes limited compared to the increasing number of LRAs. 
The Multi-Node algorithms are well suited to address these types of problems.
They operate with a fixed value $n$ for the number of nodes, given as part of the input.
In the implementation described in \S\ref{sec:solution}, a Multi-Node algorithm declares a failure if not all LRAs are allocated to the pool of $n$ nodes.
However, the LRA allocation, available after the algorithm terminates, is an appropriate solution for the problem with a given node value $n$.
Depending on the optimization criterion, one may decide to adopt the Multi-Node Algorithms with Replica Spreading, if the number of accepted LRAs is to be maximized, or the Multi-Node Algorithm with Application-Node Matching if the node utilization is to be maximized. 

\section{Related Work}
\label{sec:relatedwork}

\myparagraph{Cluster management} Resource management systems in shared clusters can be divided into two categories: centralized and decentralized systems.
Centralized approaches assign resources based on user requests~\cite{zhang2014fuxi,verma2015large,vavilapalli2013apache} or framework offers~\cite{hindman2011mesos}.
Multiple resources are negotiated among diverse applications through a central resource manager.
To make the procedure fair and avoid resource starvation, Dominant Resource Fairness~\cite{ghodsi2011dominant}, capacity or fair scheduling are adopted for resource sharing among multiple jobs.
Decentralized approaches~\cite{sun2018rose,lo2015heracles,boutin2014apollo,karanasos2015mercury} are developed for clusters that expect a high throughput or high cluster utilization. However, the goal of these works is to enable sub-second resource allocation and task scheduling at runtime without solving a global optimization problem with complex placement constraints. 

\myparagraph{LRA scheduling} YARN~\cite{vavilapalli2013apache} mainly supports the affinity constraints related to nodes/racks.
Borg~\cite{verma2015large} and ROSE~\cite{sun2018rose} use machine scoring mechanism for matching a specific collection of nodes to the requirements of the applications.
Graph-based approaches~\cite{isard2009quincy,gog2016firmament} model the scheduling problem as a min-cost max-flow optimization over a network.
However, they merely consider one dimension in the capacity constraint, and affinities to specific machines constraints. An attempt to incorporate those additional features in Aladdin~\cite{wu2019aladdin} makes it prohibitive for applying powerful min-cost max-flow methods. 

Application-level affinity is increasingly important. Kubernetes scheduler~\cite{k8s} is responsible for selecting the best node for each incoming pod. A pod is referred to as an independent execution unit and is equivalent to one replica of an LRA in this paper.
A \textit{ReplicaSet} parameter ensures that a specified number of pods are running anytime.
However, it considers one pod at each scheduling round and implements the node selection in a filtering phase.
The nodes that cannot run the pod are ruled out considering the specifications in the node/pod affinity. 
This design leads to one-shot resource allocation to a pod rather than considering it as a global optimization problem.

Medea~\cite{garefalakis2018medea} formulates the placement problem as an ILP and employs heuristics periodically to consider multiple LRAs at once at a lower scheduling latency.
However, the focus of the authors is on scheduling a small batch of LRAs.
By contrast, our work addresses pre-execution resource planning for the whole set of LRAs.

We also refer the reader to thorough surveys on wide-ranging bin packing algorithm design~\cite{BinPacking-Handbook2013,MartelloToth-book-Ch8,Panigrahy}. 

In addition, a huge body of machine learning and reinforcement learning based scheduling techniques offer alternatives for scheduling LRAs to mitigate the limitations of manual specification and resource estimation -- which usually require expert knowledge and operational experience -- in the process of requirement engineering.
\textsc{Lra}Sched~\cite{LRASched2021} employs online machine learning for auto-estimating the size of LRAs' containers and the degree of affinity.
Metis \cite{wang2020metis} and George \cite{li2021george} adopt deep reinforcement learning (DRL) to automatically learn to place LRAs based on observing the incurred reward and iteratively improving the scheduling policy. However, these works heavily depend on a huge number of high-quality workload logs, which are feasible for big companies but will place a huge obstacle on small businesses and academic organizations. Due to the exponential space of actions, DRL-based solutions are also limited to small-scale optimization problem, and thus only applicable to on-the-fly decision making. 

\myparagraph{Interference-aware LRA runtime management} There is a substantial body of research on interference-aware LRA scheduling and runtime management. 
Paragon~\cite{delimitrou2013paragon} and Quasar~\cite{wocao} use multi-variable statistical classifiers to predict the expected interference among co-located LRAs.
ROSEQ~\cite{yang2020performance} and Toposch~\cite{hu2020toposch} devise performance-aware scheduling mechanisms that can safely co-locate batch jobs together with LRAs through elaborately monitoring the runtime performance of the LRAs.
However, kernel/application-level counters are leveraged to track the runtime performance of the LRAs as a whole, without discussing the replicas and their impact on the scheduling quality. 
Overall, the focus of these research works prioritizes the performance guarantee through effective container isolating and low-cost preemption.
They are orthogonal to the resource provisioning scheme developed in this paper and offer supplementary mechanisms in the runtime execution stage.

\section{Conclusions and Future Work}
\label{sec:conclusion}

Resource provisioning of shared clusters is extremely important for minimizing the operating cost and ensuring that the scheduling systems meet both current and future demands.
LRA workloads add further complexity to resource provisioning since they run from hours to months, typically having time-varying resource requirements and co-location affinity constraints.
Careless or no planning often leads to poor utilization and performance of a cluster system.

This paper develops an affinity-aware resource provisioning scheme for LRA placement in shared clusters, supported by a new system model and an adjustable algorithmic toolkit.
The main benefits of that toolkit are as follows. 

\begin{itemize}[leftmargin=1.5em]
\item Consisting of dozens of algorithms with multiple parameters, there are three major approaches which complement each other.
Their implementation can be streamlined as algorithms' building blocks are of similar nature.
\item Application-Centric approach is the most popular one with researchers and practitioners.
However, one of its algorithms, Worst Fit Decreasing, is broadly overlooked in the literature and in practice.
Our experiments show that it often outperforms all other Application-Centric algorithms in terms of solution quality, and its execution time is comparable to the execution times of widely used First Fit Decreasing and Best Fit Decreasing algorithms from the same approach. Worst Fit Decreasing also outperforms the Node-Centric algorithms but at the cost of a slightly longer execution time.
\item The third and novel approach is Multi-Bin activation. While it involves multiple calls to one of the LRA allocation functions of Application-Centric and Node-Centric approaches, individual calls are relatively fast. If needed, the algorithm can be terminated earlier, still achieving improved solutions compared to the first two approaches.
\item The proposed toolkit is comprehensive and, together with new approaches, it encompasses a variety of the published algorithms, which can be classified as special cases of the Application-Centric and Node-Centric approaches.
A systematic summary of size measures and score functions, provided in this paper, makes the toolkit tunable to fit specific features of real-world scenarios.
We have illustrated how the tuning works based on an Alibaba public dataset and similar work could be conducted for any required scenario. 
\end{itemize}

In the future, we plan to investigate automatic algorithm selection from our algorithm pool and automatic tuning of the selected algorithm.
We also plan to integrate the proposed heuristics into Kubernetes to evaluate how theoretical study can navigate the runtime execution.

\section*{Acknowledgments}
This work was supported by UK EPSRC Grant (EP/T01461X/1), Turing Pilot Project and Turing PDEA Scheme funded by UK Alan Turing Institute. Experiments were undertaken on ARC4, part of the High Performance Computing facilities at the University of Leeds, UK.

\bibliographystyle{elsarticle-num}
\bibliography{mybib}

\end{document}